\documentclass[fleqn,10pt]{wlscirep}
\usepackage[utf8]{inputenc}
\usepackage[T1]{fontenc}
\usepackage{longtable}
\usepackage{tabularx}
\usepackage{multirow}

\title{Higher-order rich-club phenomenon in collaborative research grant networks}

\author[1,2]{Kazuki Nakajima}
\author[1,3]{Kazuyuki Shudo}
\author[2,4,5,*]{Naoki Masuda}

\affil[1]{Department of  Mathematical  and Computing Science, Tokyo  Institute  of Technology,  Meguro-ku, Tokyo  152-8552, Japan}
\affil[2]{Department  of  Mathematics,  State  University  of  New  York  at  Buffalo,  Buffalo  14260,  USA}
\affil[3]{Academic Center for Computing and Media Studies, Kyoto University, Sakyo-ku, Kyoto 606-8501, Japan}
\affil[4]{Computational  and Data-Enabled  Science  and  Engineering  Program,  State  University  of  New  York  at  Buffalo,  Buffalo  14260, USA}
\affil[5]{Faculty  of  Science  and  Engineering,  Waseda  University,  Tokyo  169-8555,  Japan}
\affil[*]{Corresponding author: naokimas@buffalo.edu}

\begin{abstract}
Modern scientific work, including writing papers and submitting research grant proposals, increasingly involves researchers from different institutions.
In grant collaborations, it is known that institutions involved in many collaborations tend to densely collaborate with each other, forming rich clubs.
Here we investigate higher-order rich-club phenomena in networks of collaborative research grants among institutions and their associations with research impact.
Using publicly available data from the National Science Foundation in the US, we construct a bipartite network of institutions and collaborative grants, which distinguishes among the collaboration with different numbers of institutions.
By extending the concept and algorithms of the rich club for dyadic networks to the case of bipartite networks, we find rich clubs both in the entire bipartite network and the bipartite subnetwork induced by the collaborative grants involving a given number of institutions up to five.
We also find that the collaborative grants within rich clubs tend to be more productive in a per-dollar sense than the control.
Our results highlight advantages of collaborative grants among the institutions in the rich clubs.
\end{abstract}

\keywords{Research grants, collaboration networks, rich-club phenomenon, research impact}

\begin{document}

\flushbottom
\maketitle
%
%
\thispagestyle{empty}


\section{Introduction}
The reliance on teamwork in scientific research has increased over the last decades \cite{zeng2017, fortunato2018}.
The fraction of scientific papers written by teams of researchers and the number of authors in a scientific paper have increased over the last century on average \cite{guimera2005, wuchty2007}.
Various factors affect outcomes of scientific teamwork, including the team size (i.e., the number of authors of a paper) \cite{wuchty2007, wu2019}, internationality (i.e., the number of countries involved in a paper) \cite{hsiehchen2015, coccia2016}, ethnic diversity (i.e., the number of ethnicities involved in a paper) \cite{alshebli2018}, interdisciplinarity (i.e., the number of disciplines of authors involved in a paper) \cite{noorden2015, vincent2015}, and team freshness (i.e., fraction of authors who have not collaborated with others before) \cite{zeng2021}.
In addition, quantitative approaches to scientific collaboration networks have contributed to the understanding of patterns of collaborations among researchers \cite{newman2001_1, zeng2017} and their relations to research productivity (e.g., the number of published papers) or impact (e.g., the number of citations received by published papers) of researchers \cite{haiyan2008, ding2009, yan2009, erjia2010, abbasi2011, abbasi2012, uddin2013, ebadi2015_2, wang2016, guan2017}.

A universal trend in modern scientific teamwork is that researchers from different institutions collaborate with each other \cite{adams2005, cummings2005, jones2008}.
Such teams tend to produce papers with higher citation impacts than those written by teams confined to a single institution \cite{jones2008}. 
Patterns of co-authorships among researchers from different institutions have been characterized through analyses of collaboration networks among institutions \cite{melin1996, ye2012, chen2020}.
Grant collaboration involving multiple institutions is also a growing trend \cite{nsf_2012, nagarajan2013, ma2015}.
Ma et al.~analyzed a British collaboration network among institutions in which edges represent partnerships between two institutions in funded research projects \cite{ma2015}.
They found that universities with many edges tend to be densely connected to each other, forming a rich club.
Analyses of such grant collaboration networks may inform the government and other stakeholders on how to allocate research funding to institutions \cite{szell2015}.

In the present study, we represent collaborations among institutions on research grants as bipartite networks to investigate grant collaborations among two or more institutions.
Note that Ma et al.~investigated a dyadic collaboration network of research grants in which collaborations between three or more institutions were represented by dyadic collaborations \cite{ma2015}.
Such a projection into dyadic networks, called the one-mode projection, is a major method for analyzing networks involving higher-order interactions among nodes \cite{newman2001_1, opsahl2008, zeng2017}.
However, evidence suggests limitations of describing such higher-order data only using pairwise interactions \cite{battiston2020, torres2021}.
In fact, despite the coordination cost that collaborating institutions owe, it is not uncommon that more than two institutions participate in a funded research project \cite{adams2005, cummings2005, jonathon2007}.
Grants with large monetary amounts often require or at least encourage inter-institutional collaboration and are sometimes a main reason for collaboration among
institutions \cite{bozeman2004}.
Large grant teams in terms of the number of investigators tend to be more productive \cite{cook2015}, and collaboration with such large and productive teams tends to receive grants in the future \cite{ebadi2015}. 
These factors may also lead to an increase in the number of collaborating institutions.
Thus motivated, we investigate networks of higher-order grant collaborations among institutions.

The relationships between research funding and research productivity or impact have been investigated for individual grants \cite{lauer2016_1}, investigators \cite{defazio2009, jacob2011, beaudry2012, fortin2013, ebadi2016}, institutions \cite{mcallister1983, boyack2003, payne2003, rosenbloom2015, ma2015}, and geographical regions \cite{zucker2007}.
Understanding such relationships is expected to assist the government and other stakeholders to develop strategies for allocating research funds to different units for enhancing research productivity or impact.
Evidence supports positive correlations between the monetary amount of research funding received by an institution and its research productivity or impact \cite{mcallister1983, boyack2003, payne2003, rosenbloom2015, ma2015}.
On the other hand, the per-dollar productivity or impact of an institution that receives a large amount of research funding tends to be diminishing \cite{zhi2016, yin2018, wahls2019, aagaard2020}.
Given this, in the present study we ask the following question: do institutions participating in many collaborative grants gain advantages in their per-dollar research impact when they densely collaborate with each other (i.e., they form a rich club) in research grants?
We examine this question using bipartite-network representation of collaborative grants among institutions, which allows us to investigate relationships among rich clubs, research impact, and the collaboration size.

\begin{figure}[t]
  \begin{center}
	\includegraphics[scale=0.295]{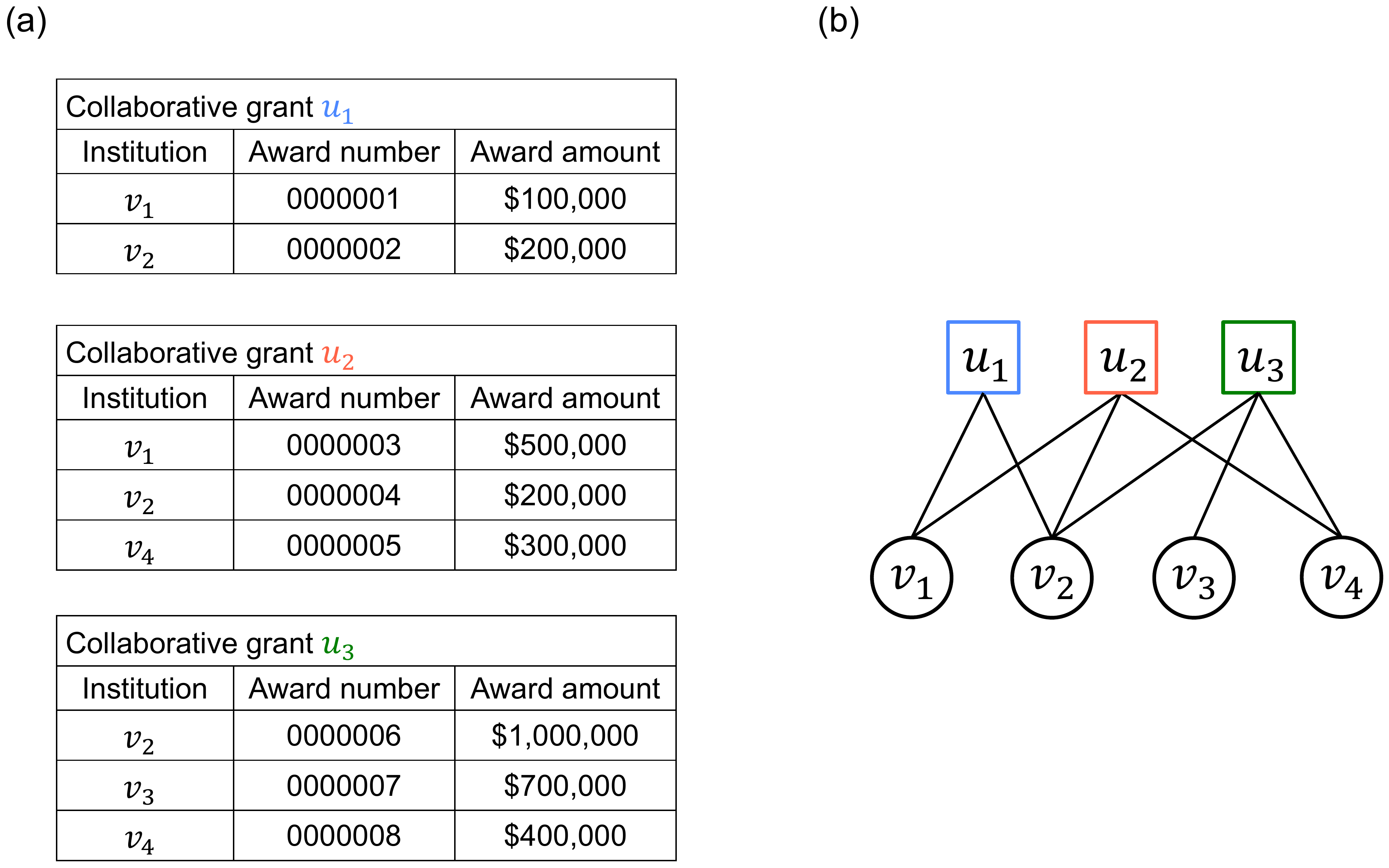}
  \end{center}
  \caption{An example of three collaborative grants and the corresponding bipartite network of institutions and collaborative grants.}
  \label{fig:1}
\end{figure}

\section{Methods}

\subsection{Construction of data sets}

\subsubsection{Collaborative grants}
We use publicly available data on the grants administered by the National Science Foundation (NSF)\footnote{\url{https://www.nsf.gov/awardsearch/download.jsp} (Accessed February 2022)}.
We focused on the collaborative grants in each of which multiple institutions participate and each institution was responsible for a separate award. 
Therefore, each collaborative grant is composed of a set of linked awards each of which is separately administered by a single institution.
For this type of collaborative grant, research proposals submitted by collaborating institutions must have the same project title beginning with `Collaborative Research:' (e.g., see the latest guide posted by the NSF\footnote{\url{https://www.nsf.gov/pubs/policydocs/pappg22_1/index.jsp} (Accessed June 2022)}. 
We confirmed that this rule was applied at least since 1999\footnote{\url{https://www.nsf.gov/pubs/1999/nsf992/cont.htm} (Accessed February 2022)}).
Therefore, we first collected the data of the awards with the project title beginning with `Collaborative Research:' and the start date between January 1, 2000 and December 31, 2020.
Second, we identified the set of institutions that received at least one such award.
Third, we used the Wikipedia APIs\footnote{\url{https://github.com/goldsmith/Wikipedia} (Accessed February 2022)} to categorize each institution into one of 48 types; see Table S1 for the complete list of institution types.
Fourth, we obtained the data of the awards received by the institutions whose type name includes `university', `college', or `school' (see Table S1 for the list of institution types that we focused on).
Among these institutions, there are 14,081 collaborative grants each of which contains at least two awards (i.e., institutions).
Fifth, for each collaborative grant, we identified the set of participating institutions, the 7-digit award number (i.e., ID) assigned to each participating institution, and the monetary amount distributed to each participating institution.

To quantify the research outputs produced under the collaborative grants, we use the Web of Science Core Collection database\footnote{\url{https://www.webofknowledge.com/} (Accessed January 2022)}.
There are 1,082,349 papers that were published between January 1, 2000 and December 31, 2020 and include at least one of the words ‘National Science Foundation' and ‘NSF' in the acknowledgment section.
The fraction of papers with acknowledgment data in this data set has increased since 2008 because the Web of Science started recording the funding acknowledgment data in August 2008\footnote{\url{http://wokinfo.com/products_tools/multidisciplinary/webofscience/fundingsearch/} (Accessed February 2022)}.
For each of these papers, we extracted the 7-digit award numbers mentioned in the acknowledgement section, the number of times cited by other papers in the database, the research disciplines assigned to the paper, which is available in the data set, the publication year, and the document type.
We retained the 1,066,324 papers whose document types are either ‘Article', ‘Review', ‘Letter', ‘Editorial Material', ‘Meeting Abstract' or ‘Proceedings Paper', as suggested in Ref.~\cite{waltman2016}.
Then, for each award comprising a collaborative grant, we identified the papers that mentioned its award number in the acknowledgment section.
We removed the collaborative grants with less than five published papers in the database because such collaborative grants often have extreme impact values due to the small number of the associated papers.
Then, we were left with 7,026 collaborative grants, each of which is associated with at least five of the 101,283 published papers.
These collaborative grants have been awarded to 570 institutions in total.

\subsubsection{Single-institution grants}

For comparison, we also analyzed the grants that were composed of just one award given to one institution.
To prepare such data, we first identified the awards of which the project title did not begin with `Collaborative Research:' and the start date was between January 1, 2000 and December 31, 2020.
There are 148,795 awards that meet these criteria and have been received by any of the 570 institutions that have participated in at least one collaborative grant.
Second, for each of these awards, we identified the institution that received the award, the 7-digit award number (i.e., ID) assigned to the institution, the monetary amount of the award, and the first and last names of a principal investigator (PI) and co-PIs.
Third, for each award, we identified the papers that mentioned its award number in the acknowledgment section.
We removed the awards associated with less than five published papers in the Web of Science database.
Then, we were left with 41,510 awards.
According to the NSF's guide, these awards belong to one of the following three types of grant: (i) single-institution grant without co-PI, (ii) single-institution grant in which all the co-PIs are from the same institution as the PI's, and (iii) collaborative grant in which at least one co-PI from a different institution from the PI's participates and the PI's institution is responsible for the award. 

We focus on the awards of types (i) and (ii) because they are genuine single-institution grants.
We found 24,866 awards of type (i) among the 41,510 awards.
It is not straightforward to classify the remaining 16,644 awards into types (ii) and (iii) because the affiliations of the co-PIs are not available in our data set.
Therefore, we attempted to identify the awards of type (ii) as follows.
First, for each co-PI in a given award, we obtain the set of candidate affiliations of the co-PI as the set of the affiliations of the authors who have the same first name initial and the same full last name as the co-PI in any of the papers associated with the award.
Second, we regard that an award is of type (ii) if and only if the set of candidate affiliations of every co-PI in the award includes the institution that has received the award.
We obtained 7,854 awards of type (ii) among the 16,644 awards with co-PIs.
Otherwise, we regard that the award is of type (iii).

In summary, we obtained $24,866 + 7,854 = 32,720$ single-institution grants, each of which is associated with at least five of the 363,116 published papers.
These grants have been awarded to 441 institutions in total.

\subsection{Bipartite network of institutions and collaborative grants}
From the data on the collaborative grants, we construct a bipartite network that consists of a set of institutions $V = \{v_1, \ldots, v_N\}$, where $N$ is the number of institutions, a set of collaborative grants $U = \{u_1, \ldots, u_M\}$, where $M$ is the number of collaborative grants, and a set of edges $E$.
An edge $(v_i, u_j)$ exists between institution $v_i$ and collaborative grant $u_j$ if and only if $v_i$ received an award in the collaborative grant $u_j$.
A unique 7-digit award number and a unique monetary amount are associated with each edge $(v_i, u_j) \in E$.
We denote by $k_i$ the degree of $v_i$, i.e., the number of awards that institution $v_i$ received from collaborative grants.
We denote by $s_j$ the degree of $u_j$, i.e., the number of collaborating institutions in collaborative grant $u_j$.
We show in Fig.~\ref{fig:1} a hypothetical bipartite network of four institutions and three collaborative grants.
In this example, we have $V = \{v_1, v_2, v_3, v_4\}$, $U = \{u_1, u_2, u_3\}$, $E = \{(v_1, u_1), (v_1, u_2), (v_2, u_1), (v_2, u_2), (v_2, u_3), (v_3, u_3), (v_4, u_2), (v_4, u_3)\}$, $k_1=2,\ k_2=3,\ k_3=1,\ k_4=2,\ s_1=2,\ s_2=3,$ and $s_3=3$.

\subsection{Detection of rich clubs}
A rich club of a dyadic network is defined as a subnetwork in which the nodes with the highest degrees (i.e., the nodes with the largest numbers of connected edges) are densely inter-connected to each other \cite{zhou2004, colizza2006}.
There are a few studies on rich clubs in bipartite networks.
Opsahl et al.~investigated rich clubs in a bipartite network of academic authors and papers \cite{opsahl2008}.
They constructed a weighted unipartite network in which the weight of each edge between two authors is equal to the number of coauthored papers, which corresponds to the one-mode projection of the bipartite network to a unipartite network, and then applied a method to detect weighted rich clubs for dyadic networks.
The same method was applied to detect a rich club in a bipartite bran network \cite{nicolas2013}, a bipartite transportation network \cite{feng2016}, and a bipartite technological network \cite{cinelli2019}.
In the present work, we investigate rich clubs in higher-order networks of collaborative grants among institutions, which one-mode projection does not characterize.
Specifically, we develop and apply a method to detect rich clubs in bipartite networks without using the one-mode projection.

We define a rich club of a given bipartite network composed of institutions and collaborative grants in which the institutions with the largest degrees densely collaborate with each other.
To compute the rich club, we first calculate the rich-club coefficient, denoted by $\phi(k)$, for the original bipartite network for a given degree $k$.
By extending the definition for dyadic networks \cite{zhou2004, colizza2006}, we define $\phi(k)$ as the number of collaborative grants that are exclusively composed of the institutions with a degree larger than $k$ divided by the maximum possible number of collaborative grants that are exclusively composed of some of these nodes.
Formally, we define
\begin{align}
\phi(k) = \frac{\lvert U_{>k} \rvert}{\sum_{i=2}^{N_{>k}}{\binom{N_{>k}}{i}}},
\label{eq:1}
\end{align}
where $U_{>k}$ is the set of collaborative grants that are exclusively composed of the institutions with a degree larger than $k$, and $N_{>k}$ is the number of institutions with a degree larger than $k$.
To examine the presence of a rich club, we need to compare $\phi(k)$ with values for a reference model \cite{colizza2006}.
Therefore, we define the normalized rich-club coefficient, denoted by $\rho(k)$, as
\begin{align}
\rho(k) = \frac{\phi(k)}{\phi_{\text{rand}}(k)},
\end{align}
where $\phi_{\text{rand}}(k)$ is the rich-club coefficient for the reference model of bipartite network.
If $\rho(k)$ is sufficiently larger than 1, we say that the institutions with a degree larger than $k$ form a rich club.
For dyadic networks, a standard choice of the reference model is the configuration model, which randomizes the edges of the original network while preserving the degree of each node \cite{colizza2006}.
Here we use a counterpart of the configuration model for bipartite networks in which we randomize the edges of the original bipartite network while preserving the degree of each institution and  each collaborative grant \cite{newman2001, nakajima2022}.
We compute $\phi_{\text{rand}}(k)$ as the rich-club coefficient averaged over 10,000 randomized bipartite networks.

\subsection{Measuring research impact for awards, institutions, and grants}

Each award in collaborative grants is associated with a monetary amount and a set of journal and conference papers supported by the award, with which we calculate the per-dollar research impact \cite{lauer2016_1} as follows.
First, to compare the citation count across different publication years and research disciplines, we normalize the number of citations received by each of the 101,283 papers, which are associated with at least one collaborative grant \cite{radicchi2008, waltman2016}.
To this end, we denote by $c$ the number of citations that a given paper $z$ has received. 
We define $c_{0}$ as the number of citations that a paper that was published in the same year as $z$ and belongs to a research discipline assigned to $z$ has received on average.
Specifically, we set $c_{0} = (\sum_{d \in D(z)} \bar{c}_{d, y(z)}) / \lvert D(z) \rvert$, where $D(z)$ is the set of the research disciplines assigned to $z$, $\lvert D(z) \rvert$ is the number of research disciplines to which $z$ belongs, $y(z)$ is the publication year of $z$, and $\bar{c}_{d, y(z)}$ is the average number of citations received by the papers published in discipline $d$ and year $y(z)$.
Each paper is assigned to at least one of the 42 research disciplines \cite{huang2020} (see Supplementary Section S2 for details).
We define the normalized number of citations received by $z$ as $c / c_{0}$.
Then, we define the per-dollar impact of the award given to institution $v_i$ in collaborative grant $u_j$, denoted by $x_{ij}$, as the sum of $c/c_0$ over all the papers associated with the award, which we then divide by the monetary amount of the award.

We measure the impact of collaborative funded research for a given subset of institutions, denoted by $V'\ (V' \subseteq V)$, as follows. 
We first calculate the average per-dollar impact of the awards in collaborative grants that the institutions in $V'$ have received, denoted by $\bar{x}_{\text{inst}}(V')$.
Then, we define the normalized impact for the set of institutions $V'$ as $\bar{x}_{\text{inst}}(V')/\bar{x}$, where $\bar{x}$ is the average per-dollar impact of all the awards in collaborative grants.
For example, when we consider the set of institutions $V'=\{v_1, v_3\}$ in a bipartite network shown in Fig.~\ref{fig:1}(b), we obtain $\bar{x}_{\text{inst}}(V') = (x_{11} + x_{12} + x_{33})/3$.
Note that $\bar{x} = (x_{11} + x_{12} + x_{21} + x_{22} + x_{23} + x_{33} + x_{42} + x_{43})/8$.
If the normalized impact is larger than 1, the impact of $V'$ is higher than the average impact of all the institutions.

We measure the impact of a given subset of collaborative grants, denoted by $U'\ (U' \subseteq U)$, as follows.
We first calculate the average per-dollar impact of the awards in $U'$, denoted by $\bar{x}_{\text{grant}}(U')$.
We are interested in whether institutional collaborations yield higher impact than the average impact of the participating institutions.
Therefore, we define the normalized impact of $U'$ as $\bar{x}_{\text{grant}}(U') / \bar{x}_{\text{inst}}(V'(U'))$, where $V'(U')$ is the set of institutions participating in at least one collaborative grant in $U'$. 
Note that $\bar{x}_{\text{inst}}(V'(U'))$ is the average per-dollar impact of the awards that the institutions in $V'(U')$ have received.
As an example, let us consider the set of collaborative grants $U'=\{u_1, u_2\}$ in a bipartite network shown in Fig.~\ref{fig:1}(b).
One obtains $\bar{x}_{\text{grant}}(U') = (x_{11} + x_{21} + x_{12} + x_{22} + x_{42})/5$.
Because set of institutions $V'(U')$ is $\{v_1, v_2, v_4\}$, one obtains $\bar{x}_{\text{inst}}(V'(U')) = (x_{11} + x_{12} + x_{21} + x_{22} + x_{23} + x_{42} + x_{43})/7$.
If the normalized impact is larger than 1, the impact of the collaborative grants in $U'$ is higher than the average impact of the institutions participating in a collaborative grant in $U'$.

To quantify the impact of single-institution grants, we adapt the above procedure for collaborative grants to the case of single-institution grants as follows.
First, we construct a bipartite network composed of institutions and single-institution grants. 
Second, we normalize the number of citations received by each of the 363,116 papers that are associated with at least one single-institution grant by the publication year and research discipline.
Then, we directly apply the definitions of impact in the case of bipartite networks of institutions and collaborative grants to the bipartite networks of institutions and single-institution grants.

\section{Results}
\begin{figure}[p]
  \begin{center}
	\includegraphics[scale=0.135]{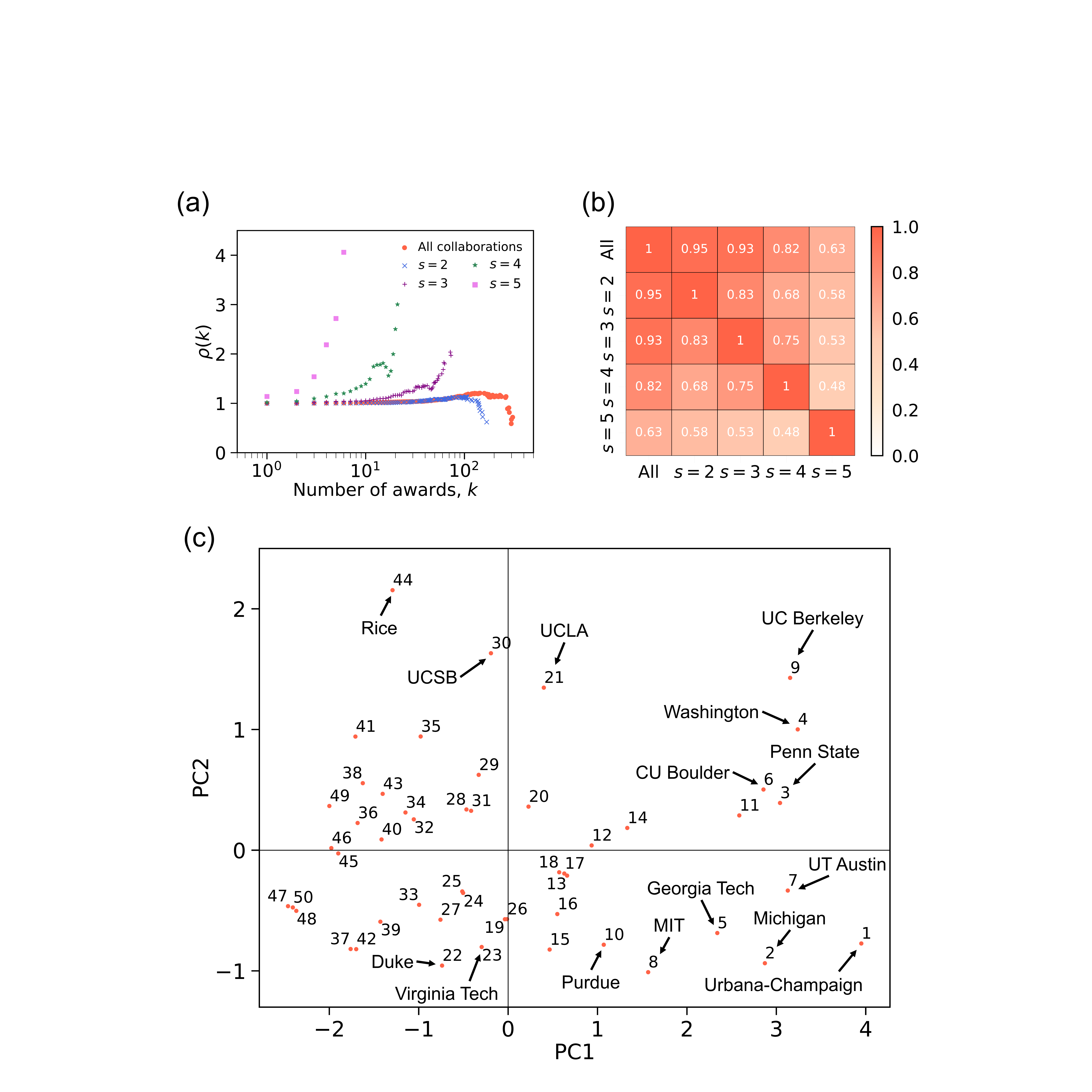}
  \end{center}
  \caption{
Rich-club phenomena in networks of grant collaboration.
(a) Normalized rich-club coefficient $\rho(k)$ as a function of the number of awards that the institution received from collaborative grants.
We measured $\rho(k)$ for the entire network (labeled ``All collaborations''), the subnetwork only composed of collaboration between $s=2$ institutions, that with $s=3$, $s=4$, and $s=5$.
In this figure, Fig.~\ref{fig:3}(b), Fig.~\ref{fig:4}(a)--(e), and Fig.~\ref{fig:5}, we omit data points for a given value of $k$ if there are less than five instances contributing to the data point.
(b) Rank correlation matrix between the different networks, where the rank is in terms of the number of awards in collaborative grants that the institution has received.
We used the top 50 institutions in the entire network to calculate the rank correlation.
(c) PCA result for the 50 institutions with the largest numbers of awards in the entire network. 
The number indicates the institution's rank in the entire network.
See the Supplementary Materials for the names of the 50 institutions.
}
  \label{fig:2}
\end{figure}

\subsection{Higher-order rich clubs in collaborative grants}

We explore possibility of higher-order rich clubs in collaborative grants.
We are also interested in how a rich-club phenomenon depends on the number of institutions in a collaborative grant.
Therefore, we calculate the normalized rich-club coefficients for the entire bipartite network and the bipartite subnetwork induced by the collaborative grants of degree (i.e., the number of collaborating institutions), $s$.
We consider $s \in \{2,3,4,5\}$ because collaborative grants with $s \ge 6$ are rare; there are less than 100 grants for each $s\ge 6$.

Figure \ref{fig:2}(a) shows the normalized rich-club coefficients for the different bipartite networks.
Figure \ref{fig:2}(a) indicates that the entire bipartite network shows a rich-club phenomenon (i.e., rich-club coefficient $> 1.10$, although this criterion is arbitrary) for the threshold of the number of awards from collaborative grants, $k$, approximately $100 \leq k \leq 200$. 
(The $P$-value is less than 0.005 for $1 \leq k \leq 193$ according to the Bonferroni-corrected permutation test; see Supplementary Section S3.)
The rich-club coefficient reaches the maximum value of approximately 1.21 at $k = 144$.
The figure also indicates that, although the bipartite subnetwork with $s=2$ has rich clubs that are statistically significant (see Supplementary Section S3), the rich-club coefficient values are modest with the largest value of 1.13. 
In contrast, the bipartite subnetwork only composed of collaborations among $s= 3$ institutions, the subnetwork restricted to $s=4$, and that restricted to $s=5$ show relatively strong and persistent rich clubs across a range of $k$. 
Therefore, the institutions that receive the largest numbers of awards from either the triadic, quartic, and quintic collaborative grants tend to more densely collaborate with each other than the institutions with the largest numbers of awards from dyadic collaborative grants.
Note that the normalized rich-club coefficient for the entire bipartite network (diamonds in Fig.~\ref{fig:2}(a)) is mostly determined by that for the subnetwork induced by the dyadic collaborative grants (crosses in Fig.~\ref{fig:2}(a)).
This is because dyadic collaborative grants are dominant in number; they account for approximately 67\% of all the collaborative grants.

We next compare the rich clubs in the different subnetworks.
We focus on the 50 institutions with the largest numbers of awards in the entire bipartite network of collaborative grants.
For these institutions, we calculate the Spearman's rank correlation coefficient in terms of the number of awards between each pair of the five bipartite networks (i.e., the entire network, $s=2$ subnetwork, $s=3$ subnetwork, $s=4$ subnetwork, and $s=5$ subnetwork).
We show the rank correlation for all pairs of networks in Fig.~\ref{fig:2}(b).
We find that the entire network is the most strongly correlated with the $s=2$ subnetwork. 
This result is expected because the collaborations between $s=2$ institutions are by far the largest contributor to the entire network.
Figure \ref{fig:2}(b) also indicates that the correlation is larger when $s$ is closer between two subnetworks.

This result led us to hypothesize that some institutions are good at securing collaborative grants involving fewer institutions, while other institutions are the opposite.
To test this hypothesis, we classify the same 50 institutions using a principal component analysis (PCA).
To run the PCA, we encode each institution into a four-dimensional vector composed of the normalized number of awards in collaborative grants with $s=2$, $s=3$, $s=4$, and $s=5$.
Specifically, we scale each entry of the vector to have mean 0 and standard deviation 1.
Then, we run the PCA on the normalized vectors using the scikit-learn library \cite{pedregosa2011}.

We show the PCA result in Fig.~\ref{fig:2}(c).
Each data point is labeled with the institution's rank in terms of the number of awards in collaborative grants that the institution has received; see Table S2 for the names of the 50 institutions.
The first two principal components, denoted by PC1 and PC2, explain 74.7\% and 13.1\% of the variance of the data, respectively.
Therefore, we conclude that the two-dimensional representation of the institutions shown in Fig.~\ref{fig:2}(c), where the two axes correspond to PC1 and PC2, is sufficient.
The eigenvector corresponding to PC1 is $(0.53, 0.54, 0.49, 0.44)$, which indicates that the number of awards from collaborative grants of any size of collaboration approximately equally contributes to PC1.
As expected, institutions with a higher rank (i.e., data points labeled with a smaller number in Fig.~\ref{fig:2}(c)) tend to have a higher PC1 value.
The eigenvector corresponding to PC2 is $(-0.25, -0.28, -0.22, 0.89)$.
Therefore, the PC2 classifies the 50 institutions into those frequent in collaborations with smaller numbers of institutions (i.e., $2 \le s\le 4$) and those frequent in collaborative grants with $s=5$. 
For example, the University of California, Berkeley ranks the 11th, 11th, 3rd, and 1st in the $s=2$, $s=3$, $s=4$, and $s=5$ subnetworks, respectively; University of Washington ranks the 6th, 2nd, 9th, and 2nd in the same four subnetworks; University of Colorado at Boulder ranks the 8th, 7th, 4th, and 4th; University of California, Los Angeles ranks the 24th, 29th, 22nd, and 7th; University of California, Santa Barbara ranks the 22nd, 38th, 42nd, and 8th; Rice University ranks the 45th, 44th, 82nd, and 6th.
The latter three universities have a much higher rank in the subnetwork with $s=5$ than that in the entire network.
The behavior of institutions with a low PC2 value is the opposite. 
For example, University of Illinois at Urbana-Champaign ranks the 1st, 1st, 8th, and 10th in the $s=2$, $s=3$, $s=4$, and $s=5$ subnetworks, respectively; University of Michigan, Ann Arbor ranks the 3rd, 3rd, 5th, and 17th in the same four subnetworks; Massachusetts Institute of Technology ranks 5th, 9th, 12th, and 28th; Duke University ranks 18th, 18th, 34th, and 55th; Virginia Polytechnic Institute and State University ranks 32nd, 19th, 14th, and 53rd.

\subsection{Research impact of the institutions with the largest numbers of collaborative grants}

\begin{figure}
  \begin{center}
	\includegraphics[scale=0.145]{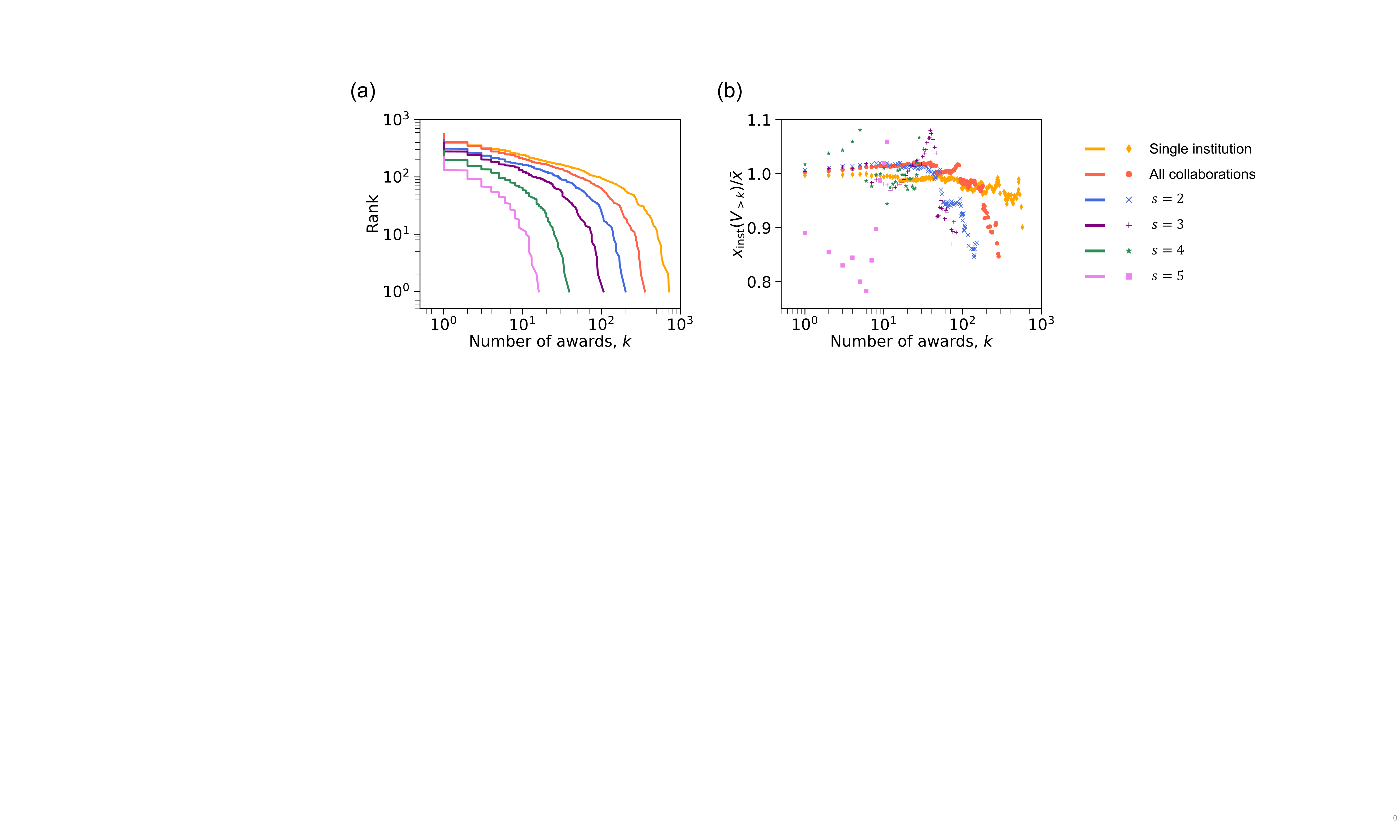}
  \end{center}
  \caption{
Research impact of award-rich institutions. 
We analyze the single-institution grants, all the collaborative grants, and the collaborative grants with different values of $s$.
(a) Rank plot of the institutions in terms of the number of awards.
(b) Normalized impact of the institutions with more than $k$ awards from grants.
We denote by $V_{>k}$ the set of those institutions.
}
  \label{fig:3}
\end{figure}

We now investigate research impact of the institutions with the largest numbers of awards from collaborative grants.
Note that these institutions form putative rich clubs.
For comparison, we also analyze the research impact of the institutions with the largest numbers of awards from single-institution grants. 
Here we analyze the data separately for all the collaborative grants, the collaborative grants comprising $s \in \{2,3,4,5\}$ institutions, and single-institution grants.

First, we show the rank plot of the number of awards received by the institution, $k$, in Fig.~\ref{fig:3}(a).
The figure indicates that $k$ is skewed toward the top-ranked institutions.
For example, the top 20\% of institutions obtained approximately 82\% of the awards in collaborative grants and approximately 79\% of the awards in single-institution grants.
This result is consistent with the concentration of research funding in top-ranked institutions observed in the NSF \cite{xie2014}, the National Institutes of Health grants in the US \cite{wahls2019, lauer2021}, and the Engineering and Physical Sciences Research Council grants in the UK \cite{ma2015}.
We also found that the top-ranked institutions less dominate the distribution of awards in the case of collaboration with a larger number of institutions (i.e., larger $s$).
For example, the top 20\% of institutions account for approximately 79\% of the awards in single-institution grants (i.e., $s=1$), 76\% for $s=2$, 70\% for $s = 3$, 60\% for $s = 4$, and 53\% for $s = 5$.
To be further quantitative, we have calculated the coefficient of variation for the distribution of the number of awards, which is equal to 1.75, 1.67, 1.49, 1.17, and 0.95 for $s=1$, $s=2$, $s=3$, $s=4$, and $s=5$, respectively;
the Gini coefficient is 0.74, 0.72, 0.66, 0.56, and 0.46 for $s=1$, $s=2$, $s=3$, $s=4$, and $s=5$, respectively.

Second, we show the normalized impact of the institutions as a function of $k$ in Fig.~\ref{fig:3}(b).
We find that the institutions with approximately 100 or more awards from collaborative grants tend to be less productive in the per-dollar sense than those with fewer awards. 
Similarly, the institutions with approximately 100 or more awards from single-institution grants tend to be less productive than those with fewer awards.
This result of the diminishing per-dollar productivity or impact at the institution level is consistent with the previous results \cite{zhi2016, yin2018, wahls2019, aagaard2020}.
Figure \ref{fig:3}(b) also indicates that similar diminishing research impact is present for collaborative grants of different collaboration sizes, $s \in \{2,3,4,5\}$.

\subsection{Research impact of the collaborative grants within rich clubs}

\begin{figure}[t]
  \begin{center}
	\includegraphics[scale=0.118]{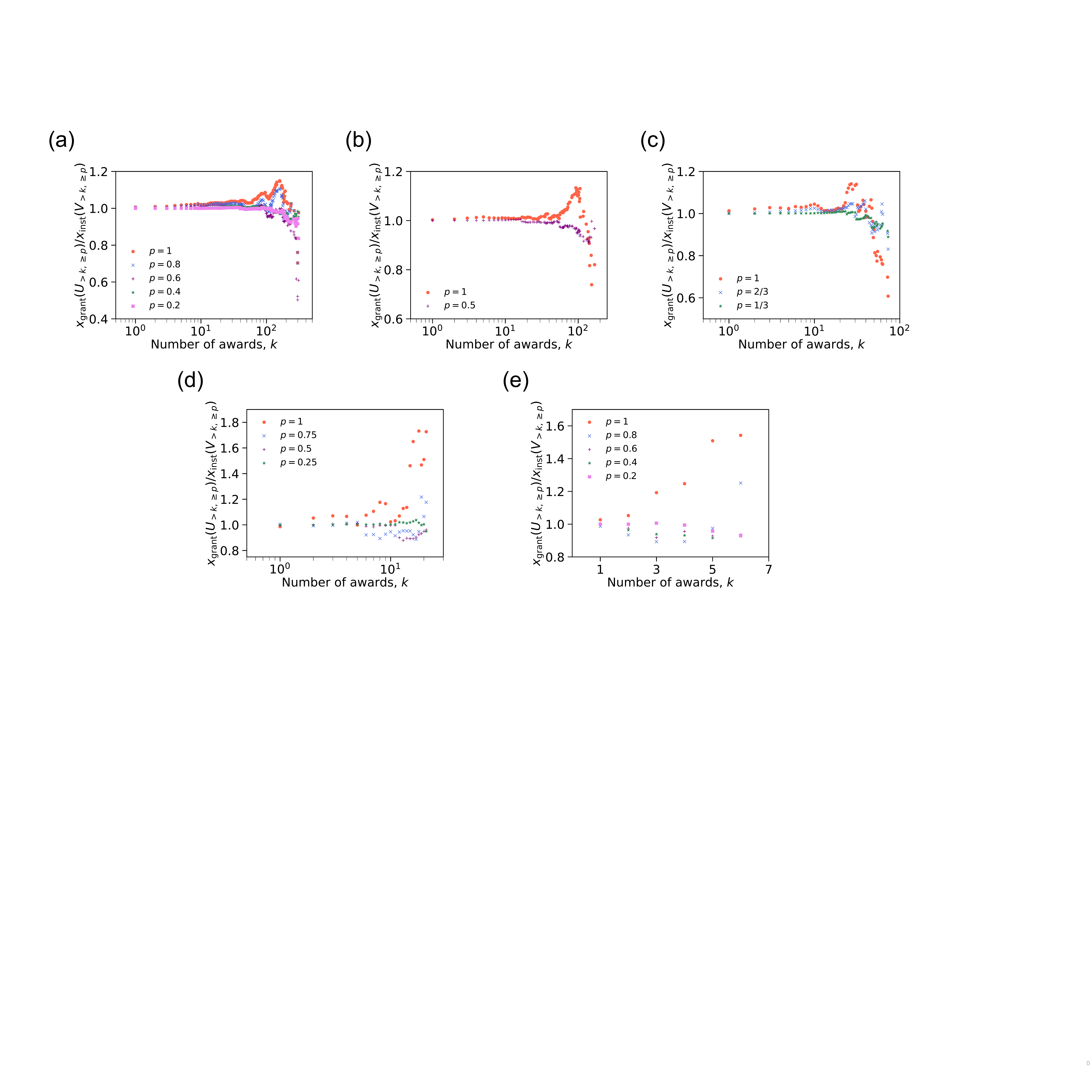}
  \end{center}
  \caption{
Advantage of collaborations among the award-rich institutions.
We plot the normalized impact of the collaborative grants in each of which fraction of the institutions receiving more than $k$ awards from collaborative grants is at least $p$.
We denote by $V_{>k, \geq p}$ the set of the institutions participating in at least one collaborative grant in $U_{>k, \geq p}$. 
(a) Entire network.
(b) Subnetwork with $s=2$. 
(c) Subnetwork with $s=3$. 
(d) Subnetwork with $s=4$. 
(e) Subnetwork with $s=5$.
}
  \label{fig:4}
\end{figure}

\begin{figure}[t]
  \begin{center}
	\includegraphics[scale=0.118]{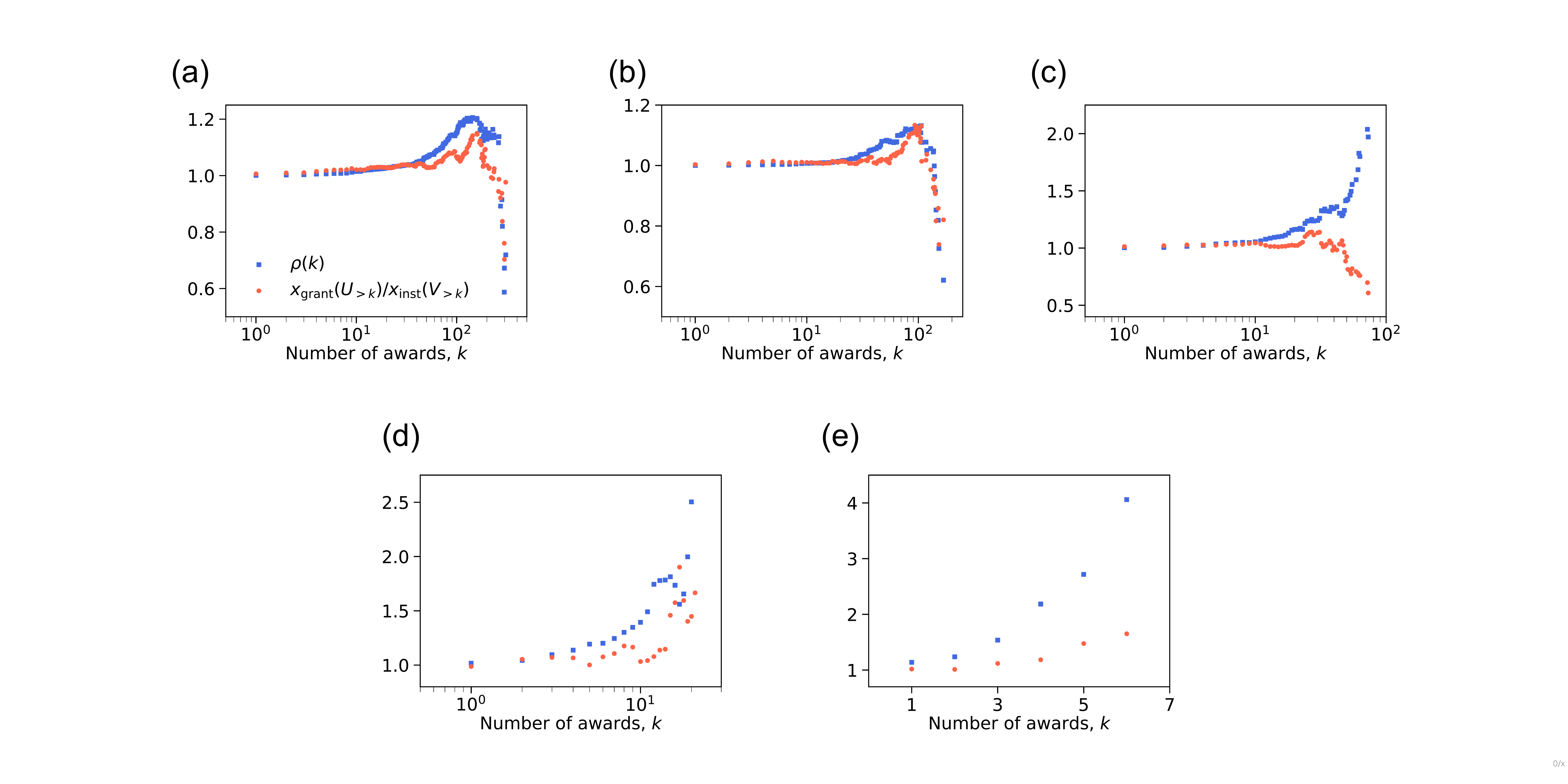}
  \end{center}
  \caption{
Overlay of the rich-club coefficient and research impact of the collaborative grants.
Each panel shows the normalized rich-club coefficient and the normalized impact as a function of the number of awards $k$ that the institution has received from collaborative grants.
(a) Entire network.
(b) Subnetwork with $s=2$. 
(c) Subnetwork with $s=3$. 
(d) Subnetwork with $s=4$. 
(e) Subnetwork with $s=5$.
}
  \label{fig:5}
\end{figure}

Given the results shown in Fig.~\ref{fig:3}, rich clubs may be detrimental to research impact because a rich club is a set of high-degree nodes, i.e., institutions with many awards. However, Fig.~\ref{fig:3} does not imply that collaborative grants among rich-club institutions are not productive; we did not look into collaboration among rich-club institutions with Fig.~\ref{fig:3}. 
Therefore, we now investigate possible associations between the rich clubs in collaborative grant networks and research impact. 
We first validate the impact of the collaborative grants within rich clubs, which are exclusively composed of the institutions with the largest numbers of awards.
We denote by $U_{>k, \geq p}$ the set of collaborative grants in which the fraction of the institutions with more than $k$ awards from collaborative grants is at least $p$.
We compare impact of the collaborative grants, $U_{>k, \geq p}$, for different $p$ values.

We show in Fig.~\ref{fig:4} the normalized impact of the collaborative grants in $U_{>k, \geq p}$ for different values of $k$ and $p$ for the entire network and the subnetwork of each collaboration size $s \in \{2, 3, 4, 5\}$. 
For the entire network, Fig.~\ref{fig:4}(a) indicates that the collaborative grants in $U_{>k, \geq p}$ with $p=1$ and large $k$ tend to be more productive than the expectation for the participating institutions.
The maximum value of the normalized impact is approximately 1.15 at $k=159$.
The figure also indicates that the collaborative grants in $U_{>k, \geq p}$ with $p=1$ for given value of $k$ tend to have a higher normalized impact than those in $U_{>k, \geq p}$ with $0 < p < 1$.
For example, at $k = 159$, the normalized impact is 1.15, 1.10, 1.00, 0.97, and 0.98 for $p=1$, $p=0.8$, $p=0.6$, $p=0.4$, and $p=0.2$, respectively.
Figures \ref{fig:4}(b)--(e) indicate that the normalized impact for $U_{>k, \geq p}$ with $p=1$ tends to be larger than 1 at large $k$ values in the subnetwork with $s \in \{2,3,4,5\}$.
This result is qualitatively the same as that for the entire collaboration network shown in Fig.~\ref{fig:4}(a).
Figures \ref{fig:4}(b)--(e) also indicate that the normalized impact for $U_{>k, \geq p}$ with $p=1$ tends to be larger than that for $U_{>k, \geq p}$ with $0 < p < 1$ in each subnetwork with $s \in \{2,3,4,5\}$.
By definition, the normalized impact of the single-institution grants is exactly equal to 1 for any $k$.
Altogether, these results indicate that collaborations among the institutions with the largest numbers of collaborative grants tend to be productive, not because such institutions tend to be strong in research but because they collaborate.

To further investigate the association between rich clubs and research impact, we investigate relationships between the normalized rich-club coefficient, $\rho(k)$, and the normalized impact of the collaborative grants that are exclusively composed of the institutions in the rich club.
We denote by $U_{>k}$ the set of collaborative grants that are exclusively composed of the institutions with more than $k$ awards from collaborative grants.
Note that $U_{>k}$ is equivalent to $U_{>k, \geq p}$ with $p=1$.
If $\rho(k)$ is sufficiently larger than 1, then $U_{>k}$ is the set of collaborative grants contained in the rich club.
Therefore, if rich clubs are associated with high research impact, the normalized impact of $U_{>k}$ should be larger than 1 for the $k$ values at which $\rho(k)$ is sufficiently larger than 1.

We show in Fig.~\ref{fig:5} the plots of $\rho(k)$ and the normalized impact of $U_{>k}$ against $k$, separately for the entire network and the subnetworks with $s \in \{2,3,4,5\}$.
The figure indicates that the normalized impact of $U_{>k}$ tends to be larger than 1 if $\rho(k)$ is larger than 1 in the entire network (Fig.~\ref{fig:5}(a)).
For example, $\rho(k)$ is largest at $k=144$.
The institutions with more than 144 awards collaborate with each other approximately 21\% more densely than in a randomized network (i.e., $\rho(144) \approx 1.21$).
The impact of the collaborative grants in $U_{>144}$ is approximately 14\% higher than expected from the average impact of the institutions participating in a collaborative grant in $U_{>144}$.
However, at $k=299$, the rich club is absent (i.e., $\rho(299) \approx 0.67$), and the impact of the collaborative grants in $U_{>299}$ is 30\% lower than the expectation for the participating institutions.
The Pearson correlation coefficient between $\rho(k)$ and the normalized impact, where we regarded a pair of these two quantities for a value of $k$ as a data point, is equal to $r = 0.85$ ($P$-value is less than 0.001).
We also found a significant positive correlation between these two quantities for the subnetwork with $s=2$ ($r = 0.89,\ P < 0.001$; see Fig.~\ref{fig:5}(b)), $s=4$ ($r = 0.61,\ P < 0.005$; see Fig.~\ref{fig:5}(d)), and $s=5$ ($r = 0.98,\ P < 0.001$; see Fig.~\ref{fig:5}(e)).
For the subnetwork with $s=3$, while we found a negative correlation ($r = -0.81,\ P < 0.001$; see Fig.~\ref{fig:5}(c)), the normalized impact tends to be larger than 1 if $\rho(k)$ is larger than 1 for approximately $1 \leq k \leq 45$.

\section{Discussion}
We investigated higher-order rich-club phenomena in networks of collaborative research grants.
To this end, we developed a method to detect rich clubs in bipartite networks.
We observed rich clubs in both the entire bipartite network and the subnetworks induced by the collaborative grants with a given number of collaborating institutions, $s$, where $s\in \{2, 3, 4, 5\}$.
The subnetworks with $s =$ 3, 4, and 5 had stronger rich clubs than that with $s=2$.
Regarding performances of rich clubs, we found that the collaborative grants within rich clubs tend to have higher per-dollar impact than the average impact expected for the institutions participating in the collaboration.
We emphasize that the higher impact of rich clubs is a genuine effect of collaboration because the impact of the single-institution grants is normalized to $1$.
These results support our hypothesis that collaborations among institutions in rich clubs are productive.

Our results extend the findings on the rich clubs in grant collaboration networks shown in a previous study \cite{ma2015} in the following two aspects.
First, we found that some collaboration-rich institutions tend to densely collaborate with each other in research grants involving fewer institutions, whereas other collaboration-rich institutions tend to do so in research grants involving more institutions.
One factor underlying this phenomenon may be strategies of individual institutions regarding interdisciplinary research projects.
Evidence suggests that interdisciplinary research projects are less likely to attract funding in a short term \cite{bromham2016}, whereas they positively contribute to long-term funding performance \cite{sun2021}.
This tendency may affect funding strategy of individual researchers and institutions, which may affect the distribution of the size of collaboration in terms of the number of institutions for the institution to which the researchers belong.
Note that Ma et al. employed the one-mode projection and therefore the impact of the size of collaboration is not a question that they focused on in their study.
Second, the benefits of rich clubs to the per-dollar research impact seem to come from collaborations among the institutions that belong to the rich clubs.
Ma et al. indicated that the rich clubs attract a large number or monetary amount of awards and tend to produce a large number of papers with high quality \cite{ma2015}.
In contrast, our results indicate that collaborations among the institutions in rich clubs are productive in terms of the per-dollar research impact, whereas the institutions themselves with many collaborations are not particularly productive.

The generality of rich clubs in grant collaboration networks deserves further investigation. 
For example, the presence of rich-club phenomena and their association with research impact may be stronger in some research disciplines than in others.
Our results do not guarantee the association between rich clubs and research impact across different disciplines.
In fact, the strength of the correlation between productivity  and institutional collaborations in writing papers substantially depends on research disciplines \cite{abramo2009}.
Rich clubs and their relevance to research impact may also depend on funding agencies.
The National Institute of Health financially encourages that multiple investigators with expertise in different health profession fields work together in research projects \cite{little2015}, which may lead to rich-club phenomena in networks in which the node is a department or institution.
Moreover, higher-order rich-club phenomena in grant collaboration networks may depend on the definition of the node.
In fact, Ma et al.~reported that a British collaboration network among investigators in which an edge represents two investigators' co-funded research projects does not have rich clubs \cite{ma2015}.

We did not address causality between rich clubs and research impact.
Furthermore, the higher impact of the collaborative grants within the rich clubs may be associated with various properties of the member institutions other than the density of their collaborations, including the internationality of the faculty \cite{mamiseishvili2009}, departmental and institutional size \cite{dundar1998}, grant type \cite{jacob2011}, and funding support from industries \cite{gulbrandsen2005}, which may affect research impact.
Additionally, there are other forms of dense mesoscopic structure of grant collaboration networks, most famous one of which is probably the community structure. 
Such other forms of dense mesoscopic structure may also affect research impact.
Examples of collaborations that may form such mesoscopic or community structures include teams composed of private universities that may be subsidized by their financial resources \cite{adams2005}, collaborations among investigators from different departmental affiliations \cite{nagarajan2013}, and collaborations between universities and industries \cite{ankrah2015}.
Moreover, many co-authorship networks among authors also show structures including the community structure and rich clubs \cite{girvan2002, opsahl2008, zeng2017}.
The present method is also applicable to the investigation of higher-order rich-club phenomena in co-authorship networks.
Further exploring the associations and causality between mesoscopic structure of networks involving higher-order interaction and research impact for various types of scientific collaborations warrants future work.

\section*{Acknowledgments}
Kazuki Nakajima was supported in part by JSPS KAKENHI Grant Number JP21J10415. 
Kazuyuki Shudo was supported in part by JSPS KAKENHI Grant Number JP21H04872. 
Naoki Masuda was supported in part by AFOSR European Office under Grant FA9550-19-1-7024, in part by Nakatani Foundation, and in part by the Japan Science and Technology Agency (JST) Moonshot R\&D (under Grant No. JPMJMS2021).


\newpage

\begin{center}
\vspace*{12pt}
{\Large Supplementary Materials for:\\
\vspace{12pt}
Higher-order rich-club phenomenon in collaborative research grant networks}
\vspace{12pt} \\
\end{center}

\newcolumntype{A}{>{\raggedright\arraybackslash}p{8cm}}
\newcolumntype{B}{>{\centering\arraybackslash}p{0.9cm}}
\newcolumntype{C}{>{\centering\arraybackslash}p{7.5cm}}
\newcolumntype{D}{>{\centering\arraybackslash}p{1.6cm}}
\newcolumntype{E}{>{\centering\arraybackslash}p{1.7cm}}

\setcounter{figure}{0}
\setcounter{table}{0}
\setcounter{section}{0}

\renewcommand{\thesection}{S\arabic{section}}
\renewcommand{\thefigure}{S\arabic{figure}}
\renewcommand{\thetable}{S\arabic{table}}

\begin{center}
\author{Kazuki Nakajima, Kazuyuki Shudo, Naoki Masuda}
\vspace{24pt} \\
\end{center}

\section{Institution types}

Table S1 shows the list of institution types that we have used.

\section{Research disciplines}

Each paper in our data set is originally assigned to at least one of the 153 research disciplines defined in the Web of Science Core Collection database.
However, some disciplines contain, for example, only one paper published in a given year.
Therefore, we used a previously proposed set of 42 disciplines that is a coarse graining of the original categorization \cite{huang2020}.
See Supplementary Table S1 in Ref. \cite{huang2020} for the mapping from the 153 disciplines to the 42 disciplines.

\section{Statistical test for normalized rich-club coefficients}

\begin{figure}[t]
  \begin{center}
	\includegraphics[scale=0.7]{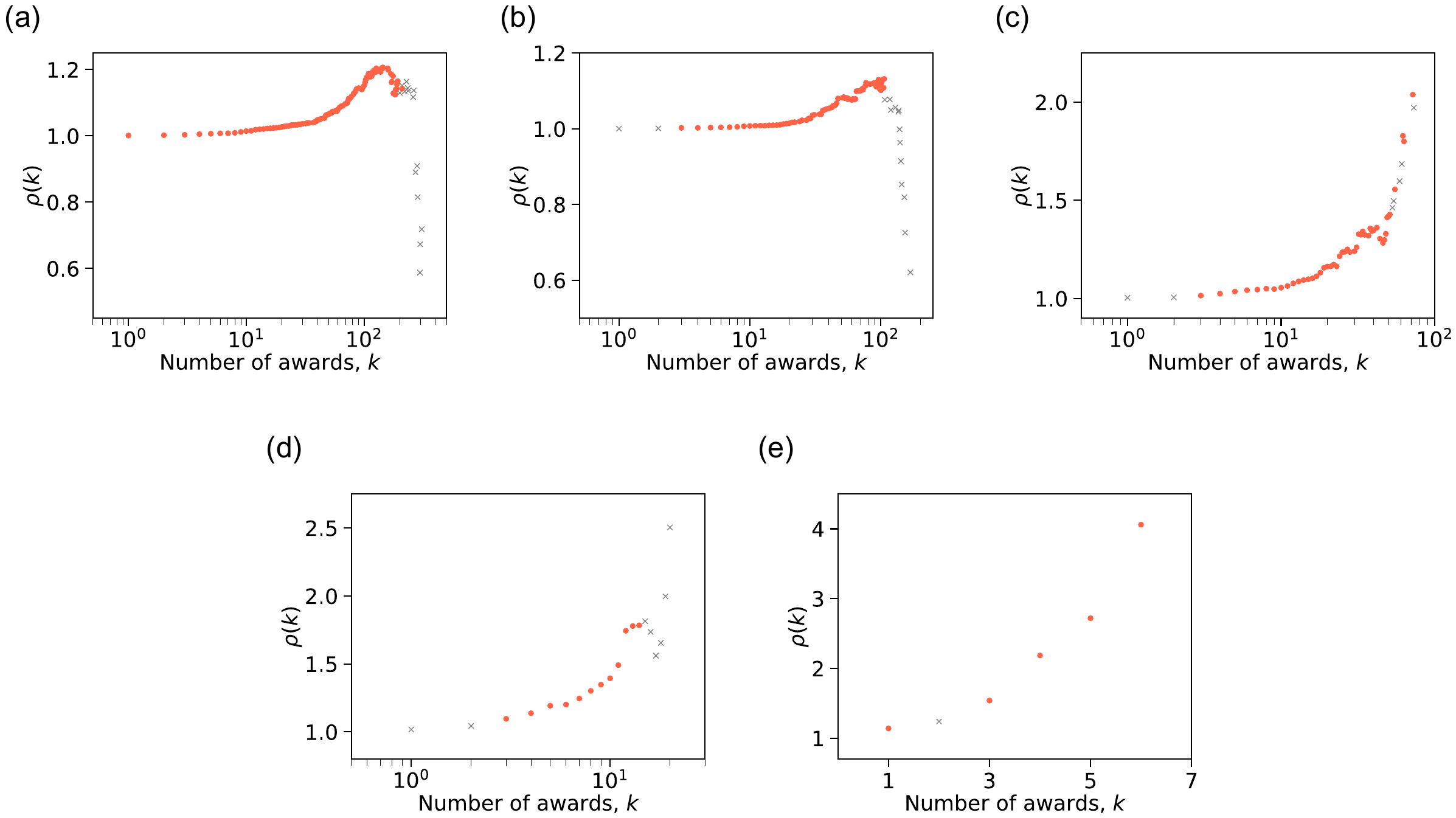}
  \end{center}
  \caption{
Normalized rich-club coefficient as a function of the number of awards received by the institution. 
A circle indicates a significant normalized rich-club coefficient ($P < 0.005$, Bonferroni-corrected permutation test).
A cross indicates a non-significant normalized rich-club coefficient.
(a) Entire network.
(b) Subnetwork with $s=2$. 
(c) Subnetwork with $s=3$. 
(d) Subnetwork with $s=4$. 
(e) Subnetwork with $s=5$.
}
  \label{fig:s1}
\end{figure}

To assess the significance of the normalized rich-club coefficient, we ran the permutation test employed in previous studies \cite{vandenheuvel2011, martijn2012, wang2019}.
We denote by $\mathcal{D}$ the set of degrees, $k$, such that there are at least five collaborative grants in which only the institutions with more than $k$ awards participate.
For a given degree $k \in \mathcal{D}$, we calculate the rich-club coefficient of the original network, i.e., $\phi(k)$, and 10,000 values of $\phi_{\text{rand}}(k)$ using the random bipartite network model.
Then, for each $k \in \mathcal{D}$, we define the $P$-value as the fraction of the $\phi_{\text{rand}}(k)$ values that are larger than $\phi(k)$.
Our null hypothesis is that $\phi(k)$ is equal to the average of the 10,000 values of $\phi_{\text{rand}}(k)$. 
The alternative hypothesis is that $\phi(k)$ is larger than the average of the 10,000 values of $\phi_{\text{rand}}(k)$.
We test the null hypothesis with Bonferroni-adjusted $\alpha$-level of $0.005/|\mathcal{D}|$ for each degree $k \in \mathcal{D}$.
We show in Fig. \ref{fig:s1} the significant and nonsignificant rich-club coefficients for the entire network and the different subnetworks.

\section{Top 50 institutions in terms of the number of collaborative grants}

Table S2 shows the top 50 institutions with the largest number of awards from collaborative grants.

\begin{center}
\newpage
\begin{longtable}{A | A}
\captionsetup{width=0.94\linewidth}
\caption{
Types of institutions. The institutions not found on the Wikipedia database are assigned `N/A' type. Those with * are the types of institution that we have used in the present study.
}
\label{table:s1}
\\
\hline
Aquarium & *Private university \\
Arboretum & Public academic health science center \\
Garden & Public agency \\
Government & *Public college \\
Health center & *Public community college \\
Hospital & *Public community college district \\
Medical center & *Public community college system \\
Military academy & *Public graduate school \\
Museum & *Public law school \\
Naval academy & *Public liberal arts college \\
*Private art and design college & *Public liberal arts university \\
*Private art and design school & *Public medical school \\
*Private college & *Public research university \\
*Private community college & *Public school of optometry \\
*Private engineering and technology school & *Public two-year college \\
*Private graduate college & *Public university \\
*Private graduate medical school & Research agency \\
*Private graduate school & Research facility \\
*Private liberal arts college & Research institute \\
*Private liberal arts university & Science center \\
*Private medical and professional school & Space agency \\
*Private medical school & Think tank \\
*Private research university & Zoo \\
*Private undergraduate and graduate school & N/A \\
\hline
\end{longtable}
\end{center}

\newpage

\begin{center}
\newpage
\begin{longtable}{| B || C | D | E |}
\captionsetup{width=0.85\linewidth}
\caption{
The top 50 institutions in terms of the number of collaborative grants.
``Public'' and ``Private'' in the last column refer to public and private research university, respectively.
}
\label{table:s1} \\
\hline
\multirow{2}{*}{Rank} & \multirow{2}{*}{Institution} & Number &  Institution \\
 & & of awards & type \\ \hline \hline 
1 & University of Illinois at Urbana-Champaign & 356 & Public \\
2 & University of Michigan, Ann Arbor & 316 & Public \\
3 & Pennsylvania State University & 308 & Public \\
4 & University of Washington & 299 & Public \\
5 & Georgia Institute of Technology & 298 & Public \\
6 & University of Colorado at Boulder & 285 & Public \\
7 & University of Texas at Austin & 282 & Public \\
8 & Massachusetts Institute of Technology & 273 & Private \\
9 & University of California, Berkeley & 273 & Public \\
10 & Purdue University & 264 & Public \\ \hline
11 & Columbia University & 261 & Private \\
12 & University of Wisconsin, Madison & 237 & Public \\
13 & Arizona State University & 235 & Public \\
14 & University of Minnesota, Twin Cities & 229 & Public \\
15 & Ohio State University & 221 & Public \\
16 & Cornell University & 211 & Private \\
17 & Stanford University & 211 & Private \\
18 & University of Arizona & 210 & Public \\
19 & Carnegie Mellon University & 201 & Private \\
20 & Oregon State University & 193 & Public \\ \hline
21 & University of California, Los Angeles & 190 & Public \\
22 & Duke University & 189 & Private \\
23 & Virginia Polytechnic Institute and State University & 185 & Public \\
24 & Rutgers University, New Brunswick & 184 & Public \\
25 & Princeton University & 183 & Private \\
26 & University of Florida & 180 & Public \\
27 & Northwestern University & 177 & Private \\
28 & University of Southern California & 177 & Private \\
29 & University of California, Davis & 176 & Public \\
30 & University of California, Santa Barbara & 172 & Public \\ \hline
31 & University of California, San Diego & 171 & Public \\
32 & University of Maryland, College Park & 169 & Public \\
33 & Harvard University & 160 & Private \\
34 & University of California, Irvine & 159 & Public \\
35 & University of California, Santa Cruz & 144 & Public \\
36 & Michigan State University & 143 & Public \\
37 & North Carolina State University & 141 & Public \\
38 & University of Massachusetts at Amherst & 138 & Public \\
39 & University of North Carolina at Chapel Hill & 138 & Public \\
40 & Yale University & 135 & Private \\ \hline
41 & University of Pennsylvania & 132 & Private \\
42 & Iowa State University & 127 & Public \\
43 & Stony Brook University & 127 & Public \\
44 & Rice University & 126 & Private \\
45 & Boston University & 123 & Private \\
46 & Johns Hopkins University & 121 & Private \\
47 & University of Pittsburgh & 121 & Public \\
48 & University of Virginia & 120 & Public \\
49 & University of Alaska Fairbanks & 118 & Public \\
50 & University of Delaware & 117 & Public \\
\hline
\end{longtable}
\end{center}

\renewcommand{\refname}{Supplementary References}


\begin{thebibliography}{10}
\urlstyle{rm}
\expandafter\ifx\csname url\endcsname\relax
  \def\url#1{\texttt{#1}}\fi
\expandafter\ifx\csname urlprefix\endcsname\relax\def\urlprefix{URL }\fi
\expandafter\ifx\csname doiprefix\endcsname\relax\def\doiprefix{DOI: }\fi
\providecommand{\bibinfo}[2]{#2}
\providecommand{\eprint}[2][]{\url{#2}}

\bibitem{zeng2017}
\bibinfo{author}{Zeng, A.} \emph{et~al.}
\newblock \bibinfo{journal}{\bibinfo{title}{The science of science: {F}rom the
  perspective of complex systems}}.
\newblock {\emph{\JournalTitle{Phys. Rep.}}}
  \textbf{\bibinfo{volume}{714--715}}, \bibinfo{pages}{1--73}
  (\bibinfo{year}{2017}).

\bibitem{fortunato2018}
\bibinfo{author}{Fortunato, S.} \emph{et~al.}
\newblock \bibinfo{journal}{\bibinfo{title}{Science of science}}.
\newblock {\emph{\JournalTitle{Science}}} \textbf{\bibinfo{volume}{359}}
  (\bibinfo{year}{2018}).
\newblock \bibinfo{note}{Article No. eaao0185}.

\bibitem{guimera2005}
\bibinfo{author}{Guimerà, R.}, \bibinfo{author}{Uzzi, B.},
  \bibinfo{author}{Spiro, J.} \& \bibinfo{author}{Amaral, L. A.~N.}
\newblock \bibinfo{journal}{\bibinfo{title}{Team assembly mechanisms determine
  collaboration network structure and team performance}}.
\newblock {\emph{\JournalTitle{Science}}} \textbf{\bibinfo{volume}{308}},
  \bibinfo{pages}{697--702} (\bibinfo{year}{2005}).

\bibitem{wuchty2007}
\bibinfo{author}{Wuchty, S.}, \bibinfo{author}{Jones, B.~F.} \&
  \bibinfo{author}{Uzzi, B.}
\newblock \bibinfo{journal}{\bibinfo{title}{The increasing dominance of teams
  in production of knowledge}}.
\newblock {\emph{\JournalTitle{Science}}} \textbf{\bibinfo{volume}{316}},
  \bibinfo{pages}{1036--1039} (\bibinfo{year}{2007}).

\bibitem{wu2019}
\bibinfo{author}{Wu, L.}, \bibinfo{author}{Wang, D.} \& \bibinfo{author}{Evans,
  J.~A.}
\newblock \bibinfo{journal}{\bibinfo{title}{Large teams develop and small teams
  disrupt science and technology}}.
\newblock {\emph{\JournalTitle{Nature}}} \textbf{\bibinfo{volume}{566}},
  \bibinfo{pages}{378--382} (\bibinfo{year}{2019}).

\bibitem{hsiehchen2015}
\bibinfo{author}{Hsiehchen, D.}, \bibinfo{author}{Espinoza, M.} \&
  \bibinfo{author}{Hsieh, A.}
\newblock \bibinfo{journal}{\bibinfo{title}{Multinational teams and
  diseconomies of scale in collaborative research}}.
\newblock {\emph{\JournalTitle{Sci. Adv.}}} \textbf{\bibinfo{volume}{1}}
  (\bibinfo{year}{2015}).
\newblock \bibinfo{note}{Article No. e1500211}.

\bibitem{coccia2016}
\bibinfo{author}{Coccia, M.} \& \bibinfo{author}{Wang, L.}
\newblock \bibinfo{journal}{\bibinfo{title}{Evolution and convergence of the
  patterns of international scientific collaboration}}.
\newblock {\emph{\JournalTitle{Proc. Natl. Acad. Sci. USA}}}
  \textbf{\bibinfo{volume}{113}}, \bibinfo{pages}{2057--2061}
  (\bibinfo{year}{2016}).

\bibitem{alshebli2018}
\bibinfo{author}{AlShebli, B.~K.}, \bibinfo{author}{Rahwan, T.} \&
  \bibinfo{author}{Woon, W.~L.}
\newblock \bibinfo{journal}{\bibinfo{title}{The preeminence of ethnic diversity
  in scientific collaboration}}.
\newblock {\emph{\JournalTitle{Nat. Commun.}}} \textbf{\bibinfo{volume}{9}}
  (\bibinfo{year}{2018}).
\newblock \bibinfo{note}{Article No. 5163}.

\bibitem{noorden2015}
\bibinfo{author}{Van~Noorden, R.}
\newblock \bibinfo{journal}{\bibinfo{title}{Interdisciplinary research by the
  numbers}}.
\newblock {\emph{\JournalTitle{Nature}}} \textbf{\bibinfo{volume}{525}},
  \bibinfo{pages}{306--307} (\bibinfo{year}{2015}).

\bibitem{vincent2015}
\bibinfo{author}{Larivière, V.}, \bibinfo{author}{Haustein, S.} \&
  \bibinfo{author}{Börner, K.}
\newblock \bibinfo{journal}{\bibinfo{title}{Long-distance interdisciplinarity
  leads to higher scientific impact}}.
\newblock {\emph{\JournalTitle{PLOS ONE}}} \textbf{\bibinfo{volume}{10}}
  (\bibinfo{year}{2015}).
\newblock \bibinfo{note}{Article No. e0122565}.

\bibitem{zeng2021}
\bibinfo{author}{Zeng, A.}, \bibinfo{author}{Fan, Y.}, \bibinfo{author}{Di,
  Z.}, \bibinfo{author}{Wang, Y.} \& \bibinfo{author}{Havlin, S.}
\newblock \bibinfo{journal}{\bibinfo{title}{Fresh teams are associated with
  original and multidisciplinary research}}.
\newblock {\emph{\JournalTitle{Nat. Hum. Behav.}}}
  \textbf{\bibinfo{volume}{5}}, \bibinfo{pages}{1314--1322}
  (\bibinfo{year}{2021}).

\bibitem{newman2001_1}
\bibinfo{author}{Newman, M. E.~J.}
\newblock \bibinfo{journal}{\bibinfo{title}{The structure of scientific
  collaboration networks}}.
\newblock {\emph{\JournalTitle{Proc. Natl. Acad. Sci. USA}}}
  \textbf{\bibinfo{volume}{98}}, \bibinfo{pages}{404--409}
  (\bibinfo{year}{2001}).

\bibitem{haiyan2008}
\bibinfo{author}{Hou, H.}, \bibinfo{author}{Kretschmer, H.} \&
  \bibinfo{author}{Liu, Z.}
\newblock \bibinfo{journal}{\bibinfo{title}{The structure of scientific
  collaboration networks in scientometrics}}.
\newblock {\emph{\JournalTitle{Scientometrics}}} \textbf{\bibinfo{volume}{75}},
  \bibinfo{pages}{189--202} (\bibinfo{year}{2008}).

\bibitem{ding2009}
\bibinfo{author}{Ding, Y.}, \bibinfo{author}{Yan, E.}, \bibinfo{author}{Frazho,
  A.} \& \bibinfo{author}{Caverlee, J.}
\newblock \bibinfo{journal}{\bibinfo{title}{Pagerank for ranking authors in
  co-citation networks}}.
\newblock {\emph{\JournalTitle{J. Assoc. Inf. Sci. Technol.}}}
  \textbf{\bibinfo{volume}{60}}, \bibinfo{pages}{2229--2243}
  (\bibinfo{year}{2009}).

\bibitem{yan2009}
\bibinfo{author}{Yan, E.} \& \bibinfo{author}{Ding, Y.}
\newblock \bibinfo{journal}{\bibinfo{title}{Applying centrality measures to
  impact analysis: A coauthorship network analysis}}.
\newblock {\emph{\JournalTitle{J. Assoc. Inf. Sci. Technol.}}}
  \textbf{\bibinfo{volume}{60}}, \bibinfo{pages}{2107--2118}
  (\bibinfo{year}{2009}).

\bibitem{erjia2010}
\bibinfo{author}{Yan, E.}, \bibinfo{author}{Ding, Y.} \& \bibinfo{author}{Zhu,
  Q.}
\newblock \bibinfo{journal}{\bibinfo{title}{Mapping library and information
  science in {C}hina: {A} coauthorship network analysis}}.
\newblock {\emph{\JournalTitle{Scientometrics}}} \textbf{\bibinfo{volume}{83}},
  \bibinfo{pages}{115--131} (\bibinfo{year}{2010}).

\bibitem{abbasi2011}
\bibinfo{author}{Abbasi, A.}, \bibinfo{author}{Altmann, J.} \&
  \bibinfo{author}{Hossain, L.}
\newblock \bibinfo{journal}{\bibinfo{title}{Identifying the effects of
  co-authorship networks on the performance of scholars: A correlation and
  regression analysis of performance measures and social network analysis
  measures}}.
\newblock {\emph{\JournalTitle{J. Informetr.}}} \textbf{\bibinfo{volume}{5}},
  \bibinfo{pages}{594--607} (\bibinfo{year}{2011}).

\bibitem{abbasi2012}
\bibinfo{author}{Abbasi, A.}, \bibinfo{author}{Chung, K. S.~K.} \&
  \bibinfo{author}{Hossain, L.}
\newblock \bibinfo{journal}{\bibinfo{title}{Egocentric analysis of
  co-authorship network structure, position and performance}}.
\newblock {\emph{\JournalTitle{Inf. Process. Manag.}}}
  \textbf{\bibinfo{volume}{48}}, \bibinfo{pages}{671--679}
  (\bibinfo{year}{2012}).

\bibitem{uddin2013}
\bibinfo{author}{Uddin, S.}, \bibinfo{author}{Hossain, L.} \&
  \bibinfo{author}{Rasmussen, K.}
\newblock \bibinfo{journal}{\bibinfo{title}{Network effects on scientific
  collaborations}}.
\newblock {\emph{\JournalTitle{PLOS ONE}}} \textbf{\bibinfo{volume}{8}}
  (\bibinfo{year}{2013}).
\newblock \bibinfo{note}{Article No. e57546}.

\bibitem{ebadi2015_2}
\bibinfo{author}{Ebadi, A.} \& \bibinfo{author}{Schiffauerova, A.}
\newblock \bibinfo{journal}{\bibinfo{title}{How to become an important player
  in scientific collaboration networks?}}
\newblock {\emph{\JournalTitle{J. Informetr.}}} \textbf{\bibinfo{volume}{9}},
  \bibinfo{pages}{809--825} (\bibinfo{year}{2015}).

\bibitem{wang2016}
\bibinfo{author}{Wang, J.}
\newblock \bibinfo{journal}{\bibinfo{title}{Knowledge creation in collaboration
  networks: {E}ffects of tie configuration}}.
\newblock {\emph{\JournalTitle{Res. Policy}}} \textbf{\bibinfo{volume}{45}},
  \bibinfo{pages}{68--80} (\bibinfo{year}{2016}).

\bibitem{guan2017}
\bibinfo{author}{Guan, J.}, \bibinfo{author}{Yan, Y.} \&
  \bibinfo{author}{Zhang, J.~J.}
\newblock \bibinfo{journal}{\bibinfo{title}{The impact of collaboration and
  knowledge networks on citations}}.
\newblock {\emph{\JournalTitle{J. Informetr.}}} \textbf{\bibinfo{volume}{11}},
  \bibinfo{pages}{407--422} (\bibinfo{year}{2017}).

\bibitem{adams2005}
\bibinfo{author}{Adams, J.~D.}, \bibinfo{author}{Black, G.~C.},
  \bibinfo{author}{Clemmons, J.~R.} \& \bibinfo{author}{Stephan, P.~E.}
\newblock \bibinfo{journal}{\bibinfo{title}{Scientific teams and institutional
  collaborations: {E}vidence from {U.S.} universities, 1981–1999}}.
\newblock {\emph{\JournalTitle{Res. Policy}}} \textbf{\bibinfo{volume}{34}},
  \bibinfo{pages}{259--285} (\bibinfo{year}{2005}).

\bibitem{cummings2005}
\bibinfo{author}{Cummings, J.~N.} \& \bibinfo{author}{Kiesler, S.}
\newblock \bibinfo{journal}{\bibinfo{title}{Collaborative research across
  disciplinary and organizational boundaries}}.
\newblock {\emph{\JournalTitle{Soc. Stud. Sci.}}}
  \textbf{\bibinfo{volume}{35}}, \bibinfo{pages}{703--722}
  (\bibinfo{year}{2005}).

\bibitem{jones2008}
\bibinfo{author}{Jones, B.~F.}, \bibinfo{author}{Wuchty, S.} \&
  \bibinfo{author}{Uzzi, B.}
\newblock \bibinfo{journal}{\bibinfo{title}{Multi-university research teams:
  Shifting impact, geography, and stratification in science}}.
\newblock {\emph{\JournalTitle{Science}}} \textbf{\bibinfo{volume}{322}},
  \bibinfo{pages}{1259--1262} (\bibinfo{year}{2008}).

\bibitem{melin1996}
\bibinfo{author}{Melin, G.} \& \bibinfo{author}{Persson, O.}
\newblock \bibinfo{journal}{\bibinfo{title}{Studying research collaboration
  using co-authorships}}.
\newblock {\emph{\JournalTitle{Scientometrics}}} \textbf{\bibinfo{volume}{36}},
  \bibinfo{pages}{363--377} (\bibinfo{year}{1996}).

\bibitem{ye2012}
\bibinfo{author}{Ye, Q.}, \bibinfo{author}{Song, H.} \& \bibinfo{author}{Li,
  T.}
\newblock \bibinfo{journal}{\bibinfo{title}{Cross-institutional collaboration
  networks in tourism and hospitality research}}.
\newblock {\emph{\JournalTitle{Tour. Manag. Perspect.}}}
  \textbf{\bibinfo{volume}{2-3}}, \bibinfo{pages}{55--64}
  (\bibinfo{year}{2012}).

\bibitem{chen2020}
\bibinfo{author}{Chen, K.}, \bibinfo{author}{Zhang, Y.}, \bibinfo{author}{Zhu,
  G.} \& \bibinfo{author}{Mu, R.}
\newblock \bibinfo{journal}{\bibinfo{title}{Do research institutes benefit from
  their network positions in research collaboration networks with industries
  or/and universities?}}
\newblock {\emph{\JournalTitle{Technovation}}}
  \textbf{\bibinfo{volume}{94--95}} (\bibinfo{year}{2020}).
\newblock \bibinfo{note}{Article No. 102002}.

\bibitem{nsf_2012}
\bibinfo{title}{National {S}cience {F}oundation. {R}esearch collaboration among
  multiple institutions is growing trend}.
\newblock
  \bibinfo{howpublished}{\url{https://www.nsf.gov/news/news_summ.jsp?cntn_id=125070}}
  (\bibinfo{year}{2012}).
\newblock \bibinfo{note}{Accessed June 2022}.

\bibitem{nagarajan2013}
\bibinfo{author}{Nagarajan, R.}, \bibinfo{author}{Kalinka, A.~T.} \&
  \bibinfo{author}{Hogan, W.~R.}
\newblock \bibinfo{journal}{\bibinfo{title}{Evidence of community structure in
  biomedical research grant collaborations}}.
\newblock {\emph{\JournalTitle{J. Biomed. Inform.}}}
  \textbf{\bibinfo{volume}{46}}, \bibinfo{pages}{40--46}
  (\bibinfo{year}{2013}).

\bibitem{ma2015}
\bibinfo{author}{Ma, A.}, \bibinfo{author}{Mondrag{\'o}n, R.~J.} \&
  \bibinfo{author}{Latora, V.}
\newblock \bibinfo{journal}{\bibinfo{title}{Anatomy of funded research in
  science}}.
\newblock {\emph{\JournalTitle{Proc. Natl. Acad. Sci. USA}}}
  \textbf{\bibinfo{volume}{112}}, \bibinfo{pages}{14760--14765}
  (\bibinfo{year}{2015}).

\bibitem{szell2015}
\bibinfo{author}{Szell, M.} \& \bibinfo{author}{Sinatra, R.}
\newblock \bibinfo{journal}{\bibinfo{title}{Research funding goes to rich
  clubs}}.
\newblock {\emph{\JournalTitle{Proc. Natl. Acad. Sci. USA}}}
  \textbf{\bibinfo{volume}{112}}, \bibinfo{pages}{14749--14750}
  (\bibinfo{year}{2015}).

\bibitem{opsahl2008}
\bibinfo{author}{Opsahl, T.}, \bibinfo{author}{Colizza, V.},
  \bibinfo{author}{Panzarasa, P.} \& \bibinfo{author}{Ramasco, J.~J.}
\newblock \bibinfo{journal}{\bibinfo{title}{Prominence and control: The
  weighted rich-club effect}}.
\newblock {\emph{\JournalTitle{Phys. Rev. Lett.}}}
  \textbf{\bibinfo{volume}{101}} (\bibinfo{year}{2008}).
\newblock \bibinfo{note}{Article No. 168702}.

\bibitem{battiston2020}
\bibinfo{author}{Battiston, F.} \emph{et~al.}
\newblock \bibinfo{journal}{\bibinfo{title}{Networks beyond pairwise
  interactions: {S}tructure and dynamics}}.
\newblock {\emph{\JournalTitle{Phys. Rep.}}} \textbf{\bibinfo{volume}{874}},
  \bibinfo{pages}{1--92} (\bibinfo{year}{2020}).

\bibitem{torres2021}
\bibinfo{author}{Torres, L.}, \bibinfo{author}{Blevins, A.~S.},
  \bibinfo{author}{Bassett, D.} \& \bibinfo{author}{Eliassi-Rad, T.}
\newblock \bibinfo{journal}{\bibinfo{title}{The why, how, and when of
  representations for complex systems}}.
\newblock {\emph{\JournalTitle{{SIAM} Rev.}}} \textbf{\bibinfo{volume}{63}},
  \bibinfo{pages}{435--485} (\bibinfo{year}{2021}).

\bibitem{jonathon2007}
\bibinfo{author}{Cummings, J.~N.} \& \bibinfo{author}{Kiesler, S.}
\newblock \bibinfo{journal}{\bibinfo{title}{Coordination costs and project
  outcomes in multi-university collaborations}}.
\newblock {\emph{\JournalTitle{Res. Policy}}} \textbf{\bibinfo{volume}{36}},
  \bibinfo{pages}{1620--1634} (\bibinfo{year}{2007}).

\bibitem{bozeman2004}
\bibinfo{author}{Bozeman, B.} \& \bibinfo{author}{Corley, E.}
\newblock \bibinfo{journal}{\bibinfo{title}{Scientists’ collaboration
  strategies: {I}mplications for scientific and technical human capital}}.
\newblock {\emph{\JournalTitle{Res. Policy}}} \textbf{\bibinfo{volume}{33}},
  \bibinfo{pages}{599--616} (\bibinfo{year}{2004}).

\bibitem{cook2015}
\bibinfo{author}{Cook, I.}, \bibinfo{author}{Grange, S.} \&
  \bibinfo{author}{Eyre-Walker, A.}
\newblock \bibinfo{journal}{\bibinfo{title}{Research groups: How big should
  they be?}}
\newblock {\emph{\JournalTitle{PeerJ}}} \textbf{\bibinfo{volume}{3}}
  (\bibinfo{year}{2015}).
\newblock \bibinfo{note}{Article No. e989}.

\bibitem{ebadi2015}
\bibinfo{author}{Ebadi, A.} \& \bibinfo{author}{Schiffauerova, A.}
\newblock \bibinfo{journal}{\bibinfo{title}{How to receive more funding for
  your research? {G}et connected to the right people!}}
\newblock {\emph{\JournalTitle{PLOS ONE}}} \textbf{\bibinfo{volume}{10}}
  (\bibinfo{year}{2015}).
\newblock \bibinfo{note}{Article No. e0133061}.

\bibitem{lauer2016_1}
\bibinfo{author}{Lauer, M.~S.}
\newblock \bibinfo{title}{Citations per dollar as a measure of productivity}.
\newblock
  \bibinfo{howpublished}{\url{https://nexus.od.nih.gov/all/2016/04/28/citations-per-dollar/}}
  (\bibinfo{year}{2016}).
\newblock \bibinfo{note}{Accessed February 2022}.

\bibitem{defazio2009}
\bibinfo{author}{Defazio, D.}, \bibinfo{author}{Lockett, A.} \&
  \bibinfo{author}{Wright, M.}
\newblock \bibinfo{journal}{\bibinfo{title}{Funding incentives, collaborative
  dynamics and scientific productivity: Evidence from the {EU} framework
  program}}.
\newblock {\emph{\JournalTitle{Res. Policy}}} \textbf{\bibinfo{volume}{38}},
  \bibinfo{pages}{293--305} (\bibinfo{year}{2009}).

\bibitem{jacob2011}
\bibinfo{author}{Jacob, B.~A.} \& \bibinfo{author}{Lefgren, L.}
\newblock \bibinfo{journal}{\bibinfo{title}{The impact of research grant
  funding on scientific productivity}}.
\newblock {\emph{\JournalTitle{J. Public Econ.}}}
  \textbf{\bibinfo{volume}{95}}, \bibinfo{pages}{1168--1177}
  (\bibinfo{year}{2011}).

\bibitem{beaudry2012}
\bibinfo{author}{Beaudry, C.} \& \bibinfo{author}{Allaoui, S.}
\newblock \bibinfo{journal}{\bibinfo{title}{Impact of public and private
  research funding on scientific production: {The} case of nanotechnology}}.
\newblock {\emph{\JournalTitle{Res. Policy}}} \textbf{\bibinfo{volume}{41}},
  \bibinfo{pages}{1589--1606} (\bibinfo{year}{2012}).

\bibitem{fortin2013}
\bibinfo{author}{Fortin, J.-M.} \& \bibinfo{author}{Currie, D.~J.}
\newblock \bibinfo{journal}{\bibinfo{title}{Big science vs. little science: How
  scientific impact scales with funding}}.
\newblock {\emph{\JournalTitle{PLOS ONE}}} \textbf{\bibinfo{volume}{8}}
  (\bibinfo{year}{2013}).
\newblock \bibinfo{note}{Article No. e65263}.

\bibitem{ebadi2016}
\bibinfo{author}{Ebadi, A.} \& \bibinfo{author}{Schiffauerova, A.}
\newblock \bibinfo{journal}{\bibinfo{title}{How to boost scientific production?
  {A} statistical analysis of research funding and other influencing factors}}.
\newblock {\emph{\JournalTitle{Scientometrics}}}
  \textbf{\bibinfo{volume}{106}}, \bibinfo{pages}{1093--1116}
  (\bibinfo{year}{2016}).

\bibitem{mcallister1983}
\bibinfo{author}{McAllister, P.~R.} \& \bibinfo{author}{Narin, F.}
\newblock \bibinfo{journal}{\bibinfo{title}{Characterization of the research
  papers of u.s. medical schools}}.
\newblock {\emph{\JournalTitle{J. Assoc. Inf. Sci. Technol.}}}
  \textbf{\bibinfo{volume}{34}}, \bibinfo{pages}{123--131}
  (\bibinfo{year}{1983}).

\bibitem{boyack2003}
\bibinfo{author}{Boyack, K.~W.} \& \bibinfo{author}{Börner, K.}
\newblock \bibinfo{journal}{\bibinfo{title}{Indicator-assisted evaluation and
  funding of research: {V}isualizing the influence of grants on the number and
  citation counts of research papers}}.
\newblock {\emph{\JournalTitle{J. Assoc. Inf. Sci. Technol.}}}
  \textbf{\bibinfo{volume}{54}}, \bibinfo{pages}{447--461}
  (\bibinfo{year}{2003}).

\bibitem{payne2003}
\bibinfo{author}{Payne, A.} \& \bibinfo{author}{Siow, A.}
\newblock \bibinfo{journal}{\bibinfo{title}{Does federal research funding
  increase university research output?}}
\newblock {\emph{\JournalTitle{The B.E. J. Econ. Anal. Policy}}}
  \textbf{\bibinfo{volume}{3}}, \bibinfo{pages}{1--24} (\bibinfo{year}{2003}).

\bibitem{rosenbloom2015}
\bibinfo{author}{Rosenbloom, J.~L.}, \bibinfo{author}{Ginther, D.~K.},
  \bibinfo{author}{Juhl, T.} \& \bibinfo{author}{Heppert, J.~A.}
\newblock \bibinfo{journal}{\bibinfo{title}{The effects of research \&
  development funding on scientific productivity: Academic chemistry,
  1990-2009}}.
\newblock {\emph{\JournalTitle{PLOS ONE}}} \textbf{\bibinfo{volume}{10}}
  (\bibinfo{year}{2015}).
\newblock \bibinfo{note}{Article No. e0138176}.

\bibitem{zucker2007}
\bibinfo{author}{Zucker, L.~G.}, \bibinfo{author}{Darby, M.~R.},
  \bibinfo{author}{Furner, J.}, \bibinfo{author}{Liu, R.~C.} \&
  \bibinfo{author}{Ma, H.}
\newblock \bibinfo{journal}{\bibinfo{title}{Minerva unbound: Knowledge stocks,
  knowledge flows and new knowledge production}}.
\newblock {\emph{\JournalTitle{Res. Policy}}} \textbf{\bibinfo{volume}{36}},
  \bibinfo{pages}{850--863} (\bibinfo{year}{2007}).

\bibitem{zhi2016}
\bibinfo{author}{Zhi, Q.} \& \bibinfo{author}{Meng, T.}
\newblock \bibinfo{journal}{\bibinfo{title}{Funding allocation, inequality, and
  scientific research output: {A}n empirical study based on the life science
  sector of natural science foundation of china}}.
\newblock {\emph{\JournalTitle{Scientometrics}}}
  \textbf{\bibinfo{volume}{106}}, \bibinfo{pages}{603--628}
  (\bibinfo{year}{2016}).

\bibitem{yin2018}
\bibinfo{author}{Yin, Z.}, \bibinfo{author}{Liang, Z.} \& \bibinfo{author}{Zhi,
  Q.}
\newblock \bibinfo{journal}{\bibinfo{title}{Does the concentration of
  scientific research funding in institutions promote knowledge output?}}
\newblock {\emph{\JournalTitle{J. Informetr.}}} \textbf{\bibinfo{volume}{12}},
  \bibinfo{pages}{1146--1159} (\bibinfo{year}{2018}).

\bibitem{wahls2019}
\bibinfo{author}{Wahls, W.~P.}
\newblock \bibinfo{journal}{\bibinfo{title}{The national institutes of health
  needs to better balance funding distributions among {US} institutions}}.
\newblock {\emph{\JournalTitle{Proc. Natl. Acad. Sci. USA}}}
  \textbf{\bibinfo{volume}{116}}, \bibinfo{pages}{13150--13154}
  (\bibinfo{year}{2019}).

\bibitem{aagaard2020}
\bibinfo{author}{Aagaard, K.}, \bibinfo{author}{Kladakis, A.} \&
  \bibinfo{author}{Nielsen, M.~W.}
\newblock \bibinfo{journal}{\bibinfo{title}{{Concentration or dispersal of
  research funding?}}}
\newblock {\emph{\JournalTitle{Quant. Sci. Stud.}}}
  \textbf{\bibinfo{volume}{1}}, \bibinfo{pages}{117--149}
  (\bibinfo{year}{2020}).

\bibitem{nsf}
\bibinfo{title}{National {S}cience {F}oundation. {D}ownload awards by year}.
\newblock
  \bibinfo{howpublished}{\url{https://www.nsf.gov/awardsearch/download.jsp}}
  (\bibinfo{year}{2022}).
\newblock \bibinfo{note}{Accessed February 2022}.

\bibitem{nsf_guide2022}
\bibinfo{title}{National {S}cience {F}oundation. {P}roposal and award policies
  and procedures guide}.
\newblock
  \bibinfo{howpublished}{\url{https://www.nsf.gov/pubs/policydocs/pappg22_1/index.jsp}}
  (\bibinfo{year}{2021}).
\newblock \bibinfo{note}{Accessed February 2022}.

\bibitem{nsf_guide1999}
\bibinfo{title}{National {S}cience {F}oundation. {T}he grant proposal guide
  ({GPG}).}
\newblock
  \bibinfo{howpublished}{\url{https://www.nsf.gov/pubs/1999/nsf992/cont.htm}}
  (\bibinfo{year}{1999}).
\newblock \bibinfo{note}{Accessed February 2022}.

\bibitem{wikipedia_api}
\bibinfo{title}{Wikipedia {P}ython library}.
\newblock \bibinfo{howpublished}{\url{https://github.com/goldsmith/Wikipedia}}.
\newblock \bibinfo{note}{Accessed February 2022}.

\bibitem{wos}
\bibinfo{title}{Web of {S}cience}.
\newblock \bibinfo{howpublished}{\url{https://www.webofknowledge.com/}}.
\newblock \bibinfo{note}{Accessed January 2022}.

\bibitem{wos_ack}
\bibinfo{title}{Web of {S}cience. {F}unding {A}cknowledgements}.
\newblock
  \bibinfo{howpublished}{\url{http://wokinfo.com/products_tools/multidisciplinary/webofscience/fundingsearch/}}.
\newblock \bibinfo{note}{Accessed February 2022}.

\bibitem{waltman2016}
\bibinfo{author}{Waltman, L.}
\newblock \bibinfo{journal}{\bibinfo{title}{A review of the literature on
  citation impact indicators}}.
\newblock {\emph{\JournalTitle{J. Informetr.}}} \textbf{\bibinfo{volume}{10}},
  \bibinfo{pages}{365--391} (\bibinfo{year}{2016}).

\bibitem{zhou2004}
\bibinfo{author}{Zhou, S.} \& \bibinfo{author}{Mondragon, R.}
\newblock \bibinfo{journal}{\bibinfo{title}{The rich-club phenomenon in the
  internet topology}}.
\newblock {\emph{\JournalTitle{IEEE Commun. Lett.}}}
  \textbf{\bibinfo{volume}{8}}, \bibinfo{pages}{180--182}
  (\bibinfo{year}{2004}).

\bibitem{colizza2006}
\bibinfo{author}{Colizza, V.}, \bibinfo{author}{Flammini, A.},
  \bibinfo{author}{Serrano, M.~A.} \& \bibinfo{author}{Vespignani, A.}
\newblock \bibinfo{journal}{\bibinfo{title}{Detecting rich-club ordering in
  complex networks}}.
\newblock {\emph{\JournalTitle{Nat. Phys.}}} \textbf{\bibinfo{volume}{2}},
  \bibinfo{pages}{110--115} (\bibinfo{year}{2006}).

\bibitem{nicolas2013}
\bibinfo{author}{Crossley, N.~A.} \emph{et~al.}
\newblock \bibinfo{journal}{\bibinfo{title}{Cognitive relevance of the
  community structure of the human brain functional coactivation network}}.
\newblock {\emph{\JournalTitle{Proc. Natl. Acad. Sci. USA}}}
  \textbf{\bibinfo{volume}{110}}, \bibinfo{pages}{11583--11588}
  (\bibinfo{year}{2013}).

\bibitem{feng2016}
\bibinfo{author}{Feng, S.}, \bibinfo{author}{Hu, B.}, \bibinfo{author}{Nie, C.}
  \& \bibinfo{author}{Shen, X.}
\newblock \bibinfo{journal}{\bibinfo{title}{Empirical study on a directed and
  weighted bus transport network in {C}hina}}.
\newblock {\emph{\JournalTitle{Physica A}}} \textbf{\bibinfo{volume}{441}},
  \bibinfo{pages}{85--92} (\bibinfo{year}{2016}).

\bibitem{cinelli2019}
\bibinfo{author}{Cinelli, M.}
\newblock \bibinfo{journal}{\bibinfo{title}{{Generalized rich-club ordering in
  networks}}}.
\newblock {\emph{\JournalTitle{J. Complex Netw.}}}
  \textbf{\bibinfo{volume}{7}}, \bibinfo{pages}{702--719}
  (\bibinfo{year}{2019}).

\bibitem{newman2001}
\bibinfo{author}{Newman, M. E.~J.}, \bibinfo{author}{Strogatz, S.~H.} \&
  \bibinfo{author}{Watts, D.~J.}
\newblock \bibinfo{journal}{\bibinfo{title}{Random graphs with arbitrary degree
  distributions and their applications}}.
\newblock {\emph{\JournalTitle{Phys. Rev. E}}} \textbf{\bibinfo{volume}{64}}
  (\bibinfo{year}{2001}).
\newblock \bibinfo{note}{Article No. 026118}.

\bibitem{nakajima2022}
\bibinfo{author}{Nakajima, K.}, \bibinfo{author}{Shudo, K.} \&
  \bibinfo{author}{Masuda, N.}
\newblock \bibinfo{journal}{\bibinfo{title}{Randomizing hypergraphs preserving
  degree correlation and local clustering}}.
\newblock {\emph{\JournalTitle{IEEE Trans. Netw. Sci. Eng.}}}
  \textbf{\bibinfo{volume}{9}}, \bibinfo{pages}{1139--1153}
  (\bibinfo{year}{2022}).

\bibitem{radicchi2008}
\bibinfo{author}{Radicchi, F.}, \bibinfo{author}{Fortunato, S.} \&
  \bibinfo{author}{Castellano, C.}
\newblock \bibinfo{journal}{\bibinfo{title}{Universality of citation
  distributions: Toward an objective measure of scientific impact}}.
\newblock {\emph{\JournalTitle{Proc. Natl. Acad. Sci. USA}}}
  \textbf{\bibinfo{volume}{105}}, \bibinfo{pages}{17268--17272}
  (\bibinfo{year}{2008}).

\bibitem{huang2020}
\bibinfo{author}{Huang, J.}, \bibinfo{author}{Gates, A.~J.},
  \bibinfo{author}{Sinatra, R.} \& \bibinfo{author}{Barabási, A.-L.}
\newblock \bibinfo{journal}{\bibinfo{title}{Historical comparison of gender
  inequality in scientific careers across countries and disciplines}}.
\newblock {\emph{\JournalTitle{Proc. Natl. Acad. Sci. USA}}}
  \textbf{\bibinfo{volume}{117}}, \bibinfo{pages}{4609--4616}
  (\bibinfo{year}{2020}).

\bibitem{pedregosa2011}
\bibinfo{author}{Pedregosa, F.} \emph{et~al.}
\newblock \bibinfo{journal}{\bibinfo{title}{Scikit-learn: {M}achine {L}earning
  in {P}ython}}.
\newblock {\emph{\JournalTitle{J. Mach. Learn. Res.}}}
  \textbf{\bibinfo{volume}{12}}, \bibinfo{pages}{2825--2830}
  (\bibinfo{year}{2011}).

\bibitem{xie2014}
\bibinfo{author}{Xie, Y.}
\newblock \bibinfo{journal}{\bibinfo{title}{``{U}ndemocracy'': inequalities in
  science}}.
\newblock {\emph{\JournalTitle{Science}}} \textbf{\bibinfo{volume}{344}},
  \bibinfo{pages}{809--810} (\bibinfo{year}{2014}).

\bibitem{lauer2021}
\bibinfo{author}{Lauer, M.~S.} \& \bibinfo{author}{Roychowdhury, D.}
\newblock \bibinfo{journal}{\bibinfo{title}{Inequalities in the distribution of
  national institutes of health research project grant funding}}.
\newblock {\emph{\JournalTitle{eLife}}} \textbf{\bibinfo{volume}{10}}
  (\bibinfo{year}{2021}).
\newblock \bibinfo{note}{Article No. e71712}.

\bibitem{bromham2016}
\bibinfo{author}{Bromham, L.}, \bibinfo{author}{Dinnage, R.} \&
  \bibinfo{author}{Hua, X.}
\newblock \bibinfo{journal}{\bibinfo{title}{Interdisciplinary research has
  consistently lower funding success}}.
\newblock {\emph{\JournalTitle{Nature}}} \textbf{\bibinfo{volume}{534}},
  \bibinfo{pages}{684--687} (\bibinfo{year}{2016}).
\newblock \bibinfo{note}{Article No. 7609}.

\bibitem{sun2021}
\bibinfo{author}{Sun, Y.}, \bibinfo{author}{Livan, G.}, \bibinfo{author}{Ma,
  A.} \& \bibinfo{author}{Latora, V.}
\newblock \bibinfo{journal}{\bibinfo{title}{Interdisciplinary researchers
  attain better long-term funding performance}}.
\newblock {\emph{\JournalTitle{Commun. Phys.}}} \textbf{\bibinfo{volume}{4}}
  (\bibinfo{year}{2021}).
\newblock \bibinfo{note}{Article No. 263}.

\bibitem{abramo2009}
\bibinfo{author}{Abramo, G.}, \bibinfo{author}{D'Angelo, C.~A.} \&
  \bibinfo{author}{Di~Costa, F.}
\newblock \bibinfo{journal}{\bibinfo{title}{Research collaboration and
  productivity: is there correlation?}}
\newblock {\emph{\JournalTitle{High. Educ.}}} \textbf{\bibinfo{volume}{57}},
  \bibinfo{pages}{155--171} (\bibinfo{year}{2009}).

\bibitem{little2015}
\bibinfo{author}{Little, M.~M.} \emph{et~al.}
\newblock \bibinfo{journal}{\bibinfo{title}{Team science as interprofessional
  collaborative research practice: a systematic review of the science of team
  science literature}}.
\newblock {\emph{\JournalTitle{J. Investig. Med.}}}
  \textbf{\bibinfo{volume}{65}}, \bibinfo{pages}{15--22}
  (\bibinfo{year}{2017}).

\bibitem{mamiseishvili2009}
\bibinfo{author}{Mamiseishvili, K.} \& \bibinfo{author}{Rosser, V.~J.}
\newblock \bibinfo{journal}{\bibinfo{title}{International and citizen faculty
  in the united states: An examination of their productivity at research
  universities}}.
\newblock {\emph{\JournalTitle{Res. High. Educ.}}}
  \textbf{\bibinfo{volume}{51}} (\bibinfo{year}{2009}).
\newblock \bibinfo{note}{Article No. 88}.

\bibitem{dundar1998}
\bibinfo{author}{Dundar, H.} \& \bibinfo{author}{Lewis, D.~R.}
\newblock \bibinfo{journal}{\bibinfo{title}{Determinants of research
  productivity in higher education}}.
\newblock {\emph{\JournalTitle{Res. High. Educ.}}}
  \textbf{\bibinfo{volume}{39}}, \bibinfo{pages}{607--631}
  (\bibinfo{year}{1998}).

\bibitem{gulbrandsen2005}
\bibinfo{author}{Gulbrandsen, M.} \& \bibinfo{author}{Smeby, J.-C.}
\newblock \bibinfo{journal}{\bibinfo{title}{Industry funding and university
  professors’ research performance}}.
\newblock {\emph{\JournalTitle{Res. Policy}}} \textbf{\bibinfo{volume}{34}},
  \bibinfo{pages}{932--950} (\bibinfo{year}{2005}).

\bibitem{ankrah2015}
\bibinfo{author}{Ankrah, S.} \& \bibinfo{author}{AL-Tabbaa, O.}
\newblock \bibinfo{journal}{\bibinfo{title}{Universities–industry
  collaboration: {A} systematic review}}.
\newblock {\emph{\JournalTitle{Scand. J. Manag.}}}
  \textbf{\bibinfo{volume}{31}}, \bibinfo{pages}{387--408}
  (\bibinfo{year}{2015}).

\bibitem{girvan2002}
\bibinfo{author}{Girvan, M.} \& \bibinfo{author}{Newman, M. E.~J.}
\newblock \bibinfo{journal}{\bibinfo{title}{Community structure in social and
  biological networks}}.
\newblock {\emph{\JournalTitle{Proc. Natl. Acad. Sci. USA}}}
  \textbf{\bibinfo{volume}{99}}, \bibinfo{pages}{7821--7826}
  (\bibinfo{year}{2002}).

\end{thebibliography}

\begin{thebibliography}{1}

\bibitem{huang2020}
J.~Huang, A.~J. Gates, R.~Sinatra, and A.-L. Barab^^c3^^a1si.
\newblock Historical comparison of gender inequality in scientific careers
  across countries and disciplines.
\newblock {\em Proc. Natl. Acad. Sci. USA}, 117:4609--4616, 2020.

\bibitem{vandenheuvel2011}
M.~P. van~den Heuvel and O.~Sporns.
\newblock Rich-club organization of the human connectome.
\newblock {\em J. Neurosci.}, 31:15775--15786, 2011.

\bibitem{martijn2012}
M.~P. van~den Heuvel, R.~S. Kahn, J.~Go^^c3^^b1i, and O.~Sporns.
\newblock High-cost, high-capacity backbone for global brain communication.
\newblock {\em Proc. Natl. Acad. Sci. USA}, 109:11372--11377, 2012.

\bibitem{wang2019}
Y.~Wang, F.~Deng, Y.~Jia, J.~Wang, S.~Zhong, H.~Huang, L.~Chen, G.~Chen, H.~Hu,
  L.~Huang, and R.~Huang.
\newblock Disrupted rich club organization and structural brain connectome in
  unmedicated bipolar disorder.
\newblock {\em Psychol. Med.}, 49:510--518, 2019.

\end{thebibliography}
\end{document}